\documentclass[prd,superscriptaddress,amsfonts,amssymb,amsmath,showpacs,twocolumn]{revtex4-2}
\usepackage{bm}
\usepackage{amsfonts}
\usepackage{latexsym}
\usepackage[latin1]{inputenc}
\usepackage{graphicx}
\usepackage{amsmath}
\usepackage{palatino}
\usepackage{mathpazo}
\usepackage{textcomp}
\usepackage{float}
\usepackage{booktabs}
\usepackage{dcolumn}
\usepackage{ragged2e}
\usepackage{hyperref}
\hypersetup{colorlinks,citecolor=blue}
\hypersetup{colorlinks=true,linkcolor=red,filecolor=magenta,    urlcolor=blue}
\usepackage{amsmath}
\usepackage{xcolor}
\usepackage{orcidlink}
\usepackage[caption=false]{subfig}
\usepackage{commath}
\def\jnl@style{\it}
\def\aaref@jnl#1{{\jnl@style#1}}

\def\aaref@jnl#1{{\jnl@style#1}}

\def\aj{\aaref@jnl{AJ}}                   % Astronomical Journal
\def\apj{\aaref@jnl{ApJ}}                 % Astrophysical Journal
\def\apjl{\aaref@jnl{ApJ}}                % Astrophysical Journal, Letters
\def\apjs{\aaref@jnl{ApJS}}               % Astrophysical Journal, Supplement
\def\apss{\aaref@jnl{Ap\&SS}}             % Astrophysics and Space Science
\def\aap{\aaref@jnl{A\&A}}                % Astronomy and Astrophysics
\def\aapr{\aaref@jnl{A\&A~Rev.}}          % Astronomy and Astrophysics Reviews
\def\aaps{\aaref@jnl{A\&AS}}              % Astronomy and Astrophysics, Supplement
\def\mnras{\aaref@jnl{Mon.~Not.~Roy.~Astron.~Soc.}}             % Monthly Notices of the RAS
\def\prd{\aaref@jnl{Phys.~Rev.~D}}        % Physical Review D
\def\plb{\aaref@jnl{Phys.~Lett.~B}}        % Physics Letters B
\def\prc{\aaref@jnl{Phys.~Rev.~C}}  % Physical Review C
\def\prl{\aaref@jnl{Phys.~Rev.~Lett.}}    % Physical Review Letters
\def\qjras{\aaref@jnl{QJRAS}}             % Quarterly Journal of the RAS
\def\skytel{\aaref@jnl{S\&T}}             % Sky and Telescope
\def\ssr{\aaref@jnl{Space~Sci.~Rev.}}     % Space Science Reviews
\def\zap{\aaref@jnl{ZAp}}                 % Zeitschrift fuer Astrophysik
\def\nat{\aaref@jnl{Nature}}              % Nature
\def\aplett{\aaref@jnl{Astrophys.~Lett.}} % Astrophysics Letters
\def\apspr{\aaref@jnl{Astrophys.~Space~Phys.~Res.}} % Astrophysics Space Physics Research
\def\physrep{\aaref@jnl{Phys.~Rep.}}      % Physics Reports
\def\physscr{\aaref@jnl{Phys.~Scr}}       % Physica Scripta
\def\commat{\aaref@jnl{Comm.~Math.~Phys.}}              % Communications in Mathematical Physics
\def\science{\aaref@jnl{Science}}               % Science
\def\cqg{\aaref@jnl{Classical Quant.~Grav.}}            % Classical and Quantum Gravity
\def\jpcs{\aaref@jnl{JPCS}}                                     % Journal of Physics Conference Series
\def\ijmpd{\aaref@jnl{Int.~J.~Mod.~Phys.~D}}                    % International Journal of Modern Physics D
\def\grg{\aaref@jnl{Gen.~Relat.~Gravit.}}               % General Relativity and Gravitation
\def\rpp{\aaref@jnl{Rep.~Prog.~Phys.}}          % Reports on Progress in Physics
\def\npa{\aaref@jnl{Nucl.~Phys.~A}}        % Nuclear Physics A
\def\lrr{\aaref@jnl{Living Rev.~Rel.}}                   % Living reviews in relativity
\def\jcap{\aaref@jnl{J.~Cosmology Astropart.~Phys.}}    % Journal of cosmology and astroparticle physics
\def\rmp{\aaref@jnl{Rev.~Mod.~Phys.}}   %Reviews of modern physics
\def\epjc{\aaref@jnl{Eur.~Phys.~J.~C}}

\allowdisplaybreaks[1]
\begin{document}
%\color{red}
\color{black}       %% For one column
\title{Structural properties of charged compact stars with color-flavour-locked quarks matter}

 \author{M. K. Jasim \orcidlink{0000-0003-0888-9935}}
 \email[Email:]{mahmoodkhalid@unizwa.edu.om}
\affiliation {Department of Mathematical and Physical Sciences, College of Arts and Sciences,\\ University of Nizwa, Nizwa, Sultanate of Oman}

\author{Anirudh Pradhan
\orcidlink{0000-0002-1932-8431}} 
\email[Email:]{ pradhan.anirudh@gmail.com}
\affiliation{Department of Mathematics, Institute of Applied Sciences and Humanities, GLA University, Mathura-281 406, Uttar
Pradesh, India}

\author{Ayan Banerjee \orcidlink{0000-0003-3422-8233}} 
\email[Email:]{ayanbanerjeemath@gmail.com}
\affiliation{Astrophysics and Cosmology Research Unit, School of Mathematics, Statistics and Computer Science, University of KwaZulu--Natal, Private Bag X54001, Durban 4000, South Africa}

\author{Takol Tangphati
\orcidlink{0000-0002-6818-8404}} 
\email[Email:]{takoltang@gmail.com}
\affiliation{Department of Physics, Faculty of Science, Chulalongkorn University, \\Bangkok 10330, Thailand}

\author{Grigoris Panotopoulos \orcidlink{0000-0002-7647-4072}} \email{grigorios.panotopoulos@ufrontera.cl}
\affiliation{Centro de Astrof{\'i}sica e Gravita{\c c}{\~a}o-CENTRA, Instituto Superior T{\'e}cnico-IST, Universidade de Lisboa-UL, Av. Rovisco Pais, 1049-001 Lisboa, Portugal}
\affiliation{Departamento de Ciencias F{\'i}sicas, Universidad de la Frontera, \\ Casilla 54-D, 4811186 Temuco, Chile}

%%%%%%%%%%%%%%%%%%%%%%%%%%%%%%%%%%%%%  DATE  %%%%%%%%%%%%%%%%%%%%%%%%%%%%%%%%%%%%

\date{\today}

\begin{abstract}
The observations of pulsars with masses close to 2$M_{\odot}$ have put strong constraints on the equation of state (EoS) of neutron-rich matter at  supranuclear densities. Moreover, the exact internal composition of those objects is largely unknown to us. Aiming to reach the 2$M_{\odot}$ limit, here we investigate the impact of electric charge on properties of compact stars assuming that the charge distribution is proportional to the mass density. The study is carried out by solving the Tolman-Oppenheimer-Volkoff (TOV) equation for a well-motivated exotic quark matter in the color-flavor-locked (CFL) phase of color superconductivity. The existence of the CFL phase may be the true ground state of hadronic matter with the possibility of the existence of a pure stable quark star (QS). Concerning the equation-of-state, we obtain structural properties of quark stars and compute the mass, the radius as well as the total electric charge of the star. We analyze the dependence of the physical properties of these QSs depending on the free parameters with special attention on mass-radius relation.  We also briefly discuss the mass vs central mass density $(M-\epsilon_c)$ relation for stability, the effect of electric charge and compactness.  Finally, our results are compared with the recent observations data on mass-radius relationship. 

\end{abstract}

\maketitle

%%%%%%%%%%%%%%%%%%%%%%%
\section{Introduction}
%%%%%%%%%%%%%%%%%%%%%%%

Compact stars including neutron stars (NSs) and quark stars (QSs) are excellent test-beds to probe their internal composition and nature of particle interactions at extremely high density regime. In turn, this would provide us with a very precise description of the internal structure, allowing an equation-of-state (EoS) (i.e. the relation between pressure and density) at nuclear and  supra-nuclear densities that can provide clues to the interactions  between elementary particles as well. Many such theoretical EoSs have been proposed, ranging from 'soft'--a mass upper limit as low as 1.5 $M_{\odot}$ \cite{Brown} to `stiff'--a higher upper mass limit near 3$M_{\odot}$ \cite{Kalogera:1996ci}.
Thus, the accurate measurement of mass-radius relation and corresponding maximum mass of compact stars is important for our understanding of the EoS of matter in such high-density situations.

Presently, we have most reliable observational constraints on the
properties of the maximum-mass NSs coming form high-precision x-ray
space missions, such as the Neutron Star Interior Composition
Explorer (NICER) \cite{Arzoumanian:2017puf,Fonseca:2016tux} and enhanced X-ray Timing and Polarimetry mission (eXTP) \cite{Zhang:2016ach} are potentially even more informative. The most recent and major breakthrough came from the first NICER simultaneous mass and radius  measurement of the millisecond pulsar PSR J0030+0451 \cite{Miller:2019cac,Riley:2019yda}.
This high-quality data presently put a very strong lower limit on the maximum mass, and have already ruled out many soft EoSs. Meanwhile, the LIGO and Virgo Collaborations have announced that the gravitational wave event
GW 190814 \cite{LIGOScientific:2020zkf} indicate the existence of a compact binary merger with a (22.2 - 24.3) $M_{\odot}$ black hole and a compact object with a mass at (2.50 - 2.67) $M_{\odot}$. The event GW 190814 provides us with an unprecedented probe into the mass gap between NSs and BHs. Thus, the  secondary component of the event GW 190814 is the most massive NS discovered to date, and this can be explained either within GR or within alternative theories of gravity, see e.g. \cite{Zhang:2020zsc,Wang:2021jfc,Tan:2020ics,Astashenok:2020qds,Astashenok:2021peo}.

On the other hand, our theoretical and  observational knowledge about NSs put a strong constraint on the theoretical models of dense nuclear matter, after the discovery of pulsars with $\sim 2$ $M_{\odot}$  including the binary millisecond pulsar J1614-2230, with mass
1.928$\pm$0.017$M_{\odot}$ \cite{Fonseca:2016tux}, and the pulsar J0348+0432 with mass 2.01 $\pm$ 0.04$M_{\odot}$ \cite{Antoniadis:2013pzd}.  The latest record is held by PSR J0740+6620, a millisecond pulsar having an updated mass of 2.08$\pm$0.07$M_{\odot}$ with its predicted radius 12.35$\pm$ 0.75 \cite{Fonseca:2021wxt} provide us a totally new avenue to probe the internal structure of NS.  Recently, two independent analyses by NICER and XMM-Newton Data found the radius to be $12.39^{+1.30}_{-0.98}$ {\rm km} \cite{Riley:2021pdl} or $13.7^{+2.6}_{-1.5}$ {\rm km} \cite{Miller:2021qha} at 68\% confidence. Additionally, a new constraint R (1.6$M_{\odot}$)$ > 10.7$ {\rm km} has been set from the analysis of the tidal deformability  obtained from GW 170817 \cite{Bauswein:2019juq}. All these measurements are more reliable than traditional spectroscopic measurements, and provide key input to analyse the NS structure and corresponding EoS.

As a consequence a lot of effort has been put forward to resolve that
problem over the last couple of decades, but the final answer is still missing up to this day. With this issue the possible existence of quark matter in compact stars has attracted the
attention for decades. Initially it was pointed out in \cite{Witten:1984rs,Bodmer:1971we} that compact stars are partially or totally made of quarks. However, according to the strange quark matter (SQM) hypothesis, this form of matter is constituted of almost equal numbers of \textit{u}, \textit{d} and \textit{s}, with the \textit{s} quark number slightly smaller due to its relatively higher static mass. It has been conjectured that SQM may be the absolute ground state of strongly interacting matter \cite{Bodmer:1971we,Itoh1970}, as its energy per baryon could be less than that of the most stable atomic nucleus, such as $^{56}$Fe and $^{62}$Ni.

As a result, there are several models used to approach the strange matter hypothesis, and the MIT Bag Model is one of the most successful ones for quark confinement. Based on this model several authors have shown the stability of strange matter \cite{Farhi:1984,Madsen:1993iw,Madsen:1998uh},
which processes many desirable features inspired by Quantum Chromodynamics (QCD) and relativity. On the other hand,  the present knowledge of QCD suggests that quark matter might be in different color superconducting phases where the temperature is low and the density is high enough, 
see Ref. \cite{Lugones:2002zd,Bogadi:2020sjy,Matsuzaki:2007kg} for details. As a result, several authors have considered an even more extreme possibility depending on the details of the quark-quark interaction, such as two-flavor color superconductor (2SC) \cite{Alford:1997zt,Anastasiou:2003ds},
the color-flavor locked (CFL) phase \cite{Alford2002,Steiner2002}, and interacting quark EoS at ultra-high densities
\cite{Becerra-Vergara:2019uzm,Banerjee:2020dad,Panotopoulos:2021sbf}.

Most of the investigations have been done under the assumption of the electric charge neutrality inside the fluid sphere. But, this point of view had been challenged by several researchers \cite{Olson1,Olson2}.
According to them the matter acquires large amounts of electric charge during the gravitational collapse  or during an accretion process onto a compact object. This point has been discussed in \cite{deDiego:2004ar,Iorio}. Moreover, the presence of electrons play a crucial role in the formation of an electric dipole layer 
on the surfaces of compact stars that lead to huge electric fields of order  $10^{18}$ V/cm \cite{alcock86:a,alcock88:a}. In general, a net electric charge on compact stars may induce intense electric fields, whose effects will add up to the internal pressure of the system. The overall repulsive force will overwhelm the gravitational one, which implies that electrically charged stars can be more stable than their neutral counterparts. In this spirit electric charge in compact stars have widely been accepted among 
researcher, see Refs. \cite{Lemos:2014lza,Arbanil:2017huq} and references therein. However, in the stellar configuration, charge can be as high as $10^{20}$ Coulomb to bring  in any change in the mass-radius relation \cite{Ray:2003gt}.  The situation is even more extreme if strange stars were made of interacting quark EoS \cite{Panotopoulos:2021cxu}.

Furthermore, we find several astrophysical model with huge amounts of
charge. In Ref. \cite{Ivanov:2002jy}, author has studied the charged perfect fluid solutions assuming a linear equation-of-state. The resulting solution brings several consequences including charged isotropic and anisotropic fluid solutions for linear or non-linear EoS
\cite{Varela:2010mf,Kumar:2018rlo,Nasim:2018ghs,Thirukkanesh:2008xc,Panotopoulos:2019wsy} as well. Also, many works have been done to explore the idea of strong electric field \cite{Arbanil:2015uoa,Malheiro:2011zz}
in the presence of strange matter followed by the MIT bag model EoS.
The most notable technique for the analysis of stability against radial perturbation has been studied for charged 
strange quark stars \cite{Arbanil:2015uoa}. Solutions of the Einstein-Maxwell system of equations for static spherically symmetric
interior spacetimes were also investigated in Refs. \cite{Komathiraj:2008em,Takisa:2013tla,Zubair:2020pvg}.

Thus, the effect of electric charge on compact objects is of great interest. In this work, we will focus on the structure properties of a charged sphere assuming the EoS for CFL strange quark matter. Obtained mass-radius relations predicted by the theory, we also  compare them with existing observational  constraints like NICER measurement of PSR J0740+6620 \cite{Fonseca:2021wxt},  PSR J0348+0432 \cite{Antoniadis:2013pzd},  EXO 1745-248 \cite{Ozel:2008kb}, PSR J0751+1807 \cite{Nice:2005fi} and PSR J0030+0451 \cite{Riley:2019yda}.  This paper is organized as follows: in Sec. \ref{sec2}, we shortly discuss the Einstein-Maxwell system of equations for static spherically symmetric spacetime. In Sec. \ref{sec3} we present
the boundary conditions and explain the method used for the numerical integration of the equations. In Sec. \ref{sec4} we discuss the EoS for CFL strange quark matter. To simplify our calculation we assume that charge density is proportional to the energy density. In Sec. \ref{sec5}, we present the numerically obtained results for quark matter stars by solving the TOV equations. Subsequently, we discuss the mass-radius relation and the stability of hydrostatic equilibrium. We close with a brief discussion of the results in Sec. \ref{sec6}. Throughout we use a system of units in which $G = c = 1$ and choose the sign conventions $(+, -, -, -)$

%%%%%%%%%%%%%%%%%%%%%%%%%%%%%%%%%%%%%%%%%%%%%%%%%%%%%%%%%%%
\section{General Relativistic Sellar Structure}\label{sec2}
%%%%%%%%%%%%%%%%%%%%%%%%%%%%%%%%%%%%%%%%%%%%%%%%%%%%%%%%%%%

To begin with the construction, let us start by considering the static spherically symmetric metric, 
which is specified by the line element  ($ds^2 = g_{\nu \mu} dx^\nu dx^\mu$, where $\nu, \mu=0,1,2,3)$,
\begin{eqnarray}
ds^2 = e^{\Phi(r)} dt^{2} -e^{\Lambda(r)}dr^{2}- r^{2} d \Omega^2 \, , \label{metr}
\end{eqnarray}
where $d \Omega^2 = d\theta^2 + \sin^2\theta d\vartheta^2 $ is the line element on the unit 2-sphere with
two unknown functions $\Phi(r)$ and $\lambda(r)$, respectively.   

The stress-energy tensor $T_{\nu}{}^{\mu}$, which in the present study is written
as $T_{\nu}{}^{\mu} = M_{\nu}{}^{\mu}+ E_{\nu}{}^{\mu}$. The sum of two terms associate with
a perfect fluid source and the electromagnetic term,  
\begin{eqnarray}\label{em}
  T_{\nu}{}^{\mu} &=& (\epsilon +P )u_{\nu} u^{\mu} + 
  P \, \delta_{\nu}{}^{ \mu}
  \nonumber \\
  &&+\frac{1}{4\pi} \left( F^{\mu l} F_{\nu l} +\frac{1}{4 \pi}
    \delta_{\nu}{} ^{\mu} F_{kl} F^{kl} \right) \, .
\end{eqnarray}
Here $\epsilon(r) $, $P(r)$ and $u^{\mu}$  are the energy density, the pressure, and
the four-velocity of the fluid, respectively.  In addition 
to this the electromagnetic field tensor $F^{\nu \mu}$ satisfy the covariant Maxwell equations 
\begin{equation} [(-g)^{1/2} F^{\nu \mu}]_{, \mu} = 4\pi j^{\nu}
  (-g)^{1/2} \, ,
  \label{ecem}
\end{equation}  
where $j^\mu$ is the $4$-current density. Since the present choice of the electromagnetic field is only due to charge, thus all components of the electromagnetic field tensor vanish, except the radial component of the electric field $F^{01}$ which satisfy  $F^{01}= -F^{10}$.
Hence, the only non-vanishing component of Maxwell equations (\ref{ecem}) is given by
\begin{equation}
  E(r) = F^{01}(r)=  \frac{1}{r^2} e^{-(\Phi + \Lambda)/2}  4\pi
    \int_{0}^{r}  r'^2 \rho_{ch} e^{ \Lambda /2} dr' \,
  , \label{comp01}
\end{equation}
where $\rho_{ch} = e^{\Phi/2}j^{0}(r)$ is the electric charge distribution inside the star. The final expression (\ref{ecem}) is given by
\begin{equation}
  \frac{d q(r)}{dr}  = 4 \pi r^2 \rho_{ch} e^{\Lambda /2} \, .
  \label{Q1}
\end{equation}
Thus, one can get the full expression for energy-momentum tensor of Eq.\ (\ref{em}), which yield 
\begin{small}
\begin{equation}
T_{\nu}{}^{\mu} =\left( \begin{array}{cccc}
 \epsilon + \frac{q^2}{8\pi r^4} & 0 & 0 & 0 \\
0 & -P + \frac{q^2}{8\pi r^4} & 0 & 0 \\
0 & 0 & -P - \frac{q^2}{8\pi r^4}  & 0 \\
0 & 0 & 0 & -P - \frac{q^2}{8\pi r^4}
\end{array} \right) , \label{TEMch}
\end{equation}
\end{small}
where the electric charge is connected to the electric field through the relation ${q(r)}/{r^2} = E(r)$. The ($tt$) and ($rr$) components of the Einstein-Maxwell equations for the metric (\ref{metr}) 
and the energy-momentum tensor (\ref{em}) are then given by
\begin{eqnarray}
  e^{-\Lambda}\left(\frac{1}{r} \frac{d\Lambda}{dr} -\frac{1}{r^{2}}\right)
  +\frac{1}{r^{2}} =  8\pi 
  \left( \epsilon + \frac{q^{2}(r)}{8\pi r^4} \right) \, ,  \label{fe1q} \\
  e^{-\Lambda}\left(\frac{1}{r}\frac{d\Phi}{dr}+\frac{1}{r^{2}}
  \right) -\frac{1}{r^{2}}=  8\pi \left( P -
    \frac{q^{2}(r)}{8\pi r^4} \right) \, . \label{fe2q}
\end{eqnarray}

Here, we introduce the gravitational mass $m(r)$ inside the sphere of radius $r$ is given by
\begin{equation}
  e^{-\Lambda(r)} \equiv 1 - \frac{2  m(r)}{ r} +\frac{ q^2(r)}{ r^2 } \, .
  \label{nord}
\end{equation}
Using the Eq. (\ref{nord}), this equation (\ref{fe1q}) gives 
 \cite{bekenstein71:a,felice95:a}
\begin{equation}
  \frac{dm}{dr} = 4\pi r^2 \epsilon
  +\frac{q}{ r}\frac{dq}{dr} \, . \label{dmel}
\end{equation}

Note that the sum of two terms on the right hand side of Eq.\ (\ref{dmel}) corresponds to the mass-energy  of the stellar matter and the mass-energy of the electric field carried the electrically charged star. Moreover, from the Bianchi identities $\nabla_{\nu} T_{\mu}^{\nu} =0$, it follows
\begin{equation}
  \frac{d\Phi}{dr} = -\frac{2}{\left(\epsilon+P\right)}\left(\frac{dP}{dr} -\frac{q}{4 \pi r^4}\frac{dq}{dr} \right) \, . \label{bic}
\end{equation}

Then, replacing (\ref{Q1}) and taking into account the expressions
(\ref{bic}) into Eq. (\ref{fe2q}), we get
\begin{eqnarray}
  \frac{dP}{dr}  & = & - \frac{2 \left( m + 4\pi r^3
      \left( P - \frac{q^{2} }{4\pi r^{4} } \right) \right)}{ r^{2}
    \left( 1 - \frac{2 m}{ r} + \frac{ q^{2}}{r^{2} } \right)}
 \ (P +\epsilon)\nonumber \\ & & +\frac{q}{4 \pi r^4}\frac{dq}{dr} \, ,
\label{TOVca}
\end{eqnarray}
which is the Tolman-Oppenheimer-Volkoff (TOV) equation for  an electrically charged fluid sphere.  For $q \rightarrow 0$ these equations reduce to the ordinary TOV equations of General Relativity for electrically neutral fluid spheres.

Thus, we have six unknown functions-- $\Phi(r)$,  $m(r)$, $q(r)$, $\epsilon(r)$, $P(r)$ and $\rho_{ch}(r)$,
for which there are four equations, (\ref{Q1}), (\ref{dmel}) , (\ref{bic}) and (\ref{TOVca}).  Therefore, the above system of equations is under-determined, and we will reduce the number of unknown functions by assuming suitable conditions for charged fluid sphere (see \cite{Arbanil:2015uoa,Arbanil:2013pua} for more details). In particular,
one may consider a relation between the pressure and the density is called the equation-of-state, which is necessary to solve the equations for the stellar structure of the star. Furthermore, a linear relation between the charge distribution and the mass density will be assumed in order to solve the system of structure equations, combined with the appropriate boundary conditions.

%%%%%%%%%%%%%%%%%%%%%%%%%%%%%%%%%%%%%%%%%%%%%%%%%%%%%%%%%%%%%%

\section{Boundary conditions and the exterior vacuum region} \label{sec3}

The main aim here is to numerically solve the modified TOV 
equations inside and outside the star simultaneously for the sought solutions. In the present case, we maintain the regularity inside the fluid sphere with the following  boundary conditions: 
\begin{eqnarray}
&& m(r=0) = 0,~~~q(r=0)=0,~~~~\epsilon(r=0)=\epsilon_c,\\ \nonumber 
&& P(r=0)= P_c,~~~e^{\Lambda(r)}|_{r=0}=1, ~~\text{and}~~P(R)|_{r=R}=0, 
\end{eqnarray}
where $r = R$ is the surface of the star with $P_c$ is the central pressure and  $\epsilon_c$ is the central energy density, respectively. As our aim is to solve the TOV equations numerically for a specific EoS, here we choose an appropriate set of boundary conditions. We start with a central density $\epsilon_c$, and a corresponding central pressure $P_c$, and then integrate towards the surface where the pressure vanishes.

On the other hand, our numerical solution should match the external solution for $r > R$. The unique solution to the Einstein-Maxwell system for $r > R$ is given by the Reissner-Nordstr\"{o}m spacetime,
\begin{eqnarray}
ds^2 = F(r) dt^2-\frac{dr^2}{F(r)} - r^{2} d \Omega^2 
\end{eqnarray}
where $F(r)= \left(1-\frac{2 M}{r}+\frac{Q^2}{ r^2}\right)$. By matching the solution we can find the total mass $M$ and the total charge $Q$  of the star. Since, the  first fundamental form at the boundary implies that $g_{tt}^{-} = g_{tt}^{+}$ and $g_{rr}^{-} = g_{rr}^{+}$, which implies 
\begin{eqnarray}
 \label{eq50}
e^{\Phi(r)}|_{r=R}= F(R) \quad \mbox{and} \quad   e^{-\Lambda(r)}|_{r=R}=F(R),
\end{eqnarray}
with other conditions are $m(R) = M$ and $q(R) = Q$, besides $P(R) = 0$.

%%%%%%%%%%%%%%%%%%%%%%%%%%%%%%%%%%%%%%%%%%%%%%%%%%%%%%%%%%%%%%%%%%%%
\section{Equations of State for Quark matter and the charge density
profile} \label{sec4}
%%%%%%%%%%%%%%%%%%%%%%%%%%%%%%%%%%%%%%%%%%%%%%%%%%%%%%%%%%%%%%%%%%%%

\subsection{Specific equations-of-state}

As mentioned before, we have two degrees of freedom. Thus, we need to specify an EoS and an assumption for the charge density, as the matter content carries a net electric charge. First, we define the standard relations between pressure and energy density through the EoS. We work here within the hypothesis that quark matter is absolutely stable and thus quark stars contain roughly equal numbers of up, down, and strange quarks. Moreover, strange quark matter (SQM) is supposed to be the most stable quantum state of the hadronic matter \cite{Witten:1984rs,Bodmer:1971we}.

On the other hand, color superconductivity in quark matter has become a compelling topic that devoted to the physics of compact star interiors. At sufficiently large densities and low temperatures
that hadrons are crushed into quark matter, there is a whole family of ``color superconducting" phases  
\cite{Alford:2001dt,Nardulli:2002ma,Reddy:2002ri}. The rich phase structure is essentially due to the quark-quark interaction. Since, the interaction is strong and attractive in many channels with the possibility of many degrees of freedom for quarks like color, flavor and spin so that various kinds of BCS pairing are possible. With this assumption several models have been predicted in existence of a large variety of color superconducting states of quark matter at ultra-high densities \cite{Lugones:2002zd,Bogadi:2020sjy,Matsuzaki:2007kg}.

%%%%%%%

In the present work we focus on the three-flavor quark matter with the particular symmetry is called the color-flavor locked matter at asymptotically high density. It is widely accepted that the CFL phase is the real ground state of QCD at asymptotically large densities. Quark matter in the CFL phase is electrically neutral, and electrons cannot be present, contrary to the case of NSs where neutrality requires the presence of electrons. When the color superconducting pairing is considered in the CFL phase,  the thermodynamic potential for electric and color charge neutral CFL quark matter to order 
$\Delta^2$ is given by \cite{alf01} 
\begin{eqnarray}
 \Omega_{CFL}&=& -{3\Delta^2 \mu^2 \over \pi^2}+{6 \over \pi^2} \int_0^{\gamma_F} p^2 (p-\mu)~dp   \nonumber \\
&+&{3 \over \pi^2} \int_0^{\gamma_F} p^2 \left(\sqrt{p^2+m_s^2}-\mu \right)~dp+B, ~~~~~~~~\label{cfl}
\end{eqnarray}
where $\mu$ is the quark chemical potential and $\Delta$ denotes the color superconducting gap parameter of the CFL phase of quark matter. The first term is the contribution of the CFL condensate to $\Omega_{CFL}$, and the next two terms coming from the thermodynamic potential of (fictional) unpaired quark matter in which all quarks that are going to pair have a common Fermi momentum $\gamma_F$, with  $\gamma_F$ chosen to
minimize the thermodynamic potential of the fictional unpaired quark matter \cite{Alford:2002rj}. The final term is the bag constant.

The common Fermi momentum $\gamma_F$ is given by 
\begin{eqnarray}
\gamma_F = \bigg[\bigg( 2\mu - \sqrt{\mu^2+{m_s^2-m_u^2 \over 3}}\bigg)^2 -m_u^2 \bigg]^{1/2}
\end{eqnarray}
where $\mu=(\mu_s+\mu_u+\mu_d)/3$ is the average quark chemical potential, and $m_s$ and $m_u$ are  strange and up quark 
masses, respectively. For massless up and down quarks we get
\begin{eqnarray}
\gamma_F &=& 2\mu - \sqrt{\mu^2+{m_s^2 \over 3}} \sim \mu-{m_s^2 \over 6\mu}.  
\end{eqnarray}
With the pairing ansatz in the CFL phase \cite{Steiner2002}
\begin{eqnarray}
n_u=n_r, ~~~~~ n_d = n_g, ~~~~~\text{and}~~~~~ n_s = n_b
\end{eqnarray}
where $n_r$, $n_g$, $n_b$ and $n_u$, $n_d$, $n_s$ are color and flavor number densities respectively. The pressure, baryon number density $n_B$ , and particle number densities are easily derived and read
\begin{eqnarray}
P= - \Omega_{CFL} ~~\text{and}~~n_u = n_d = n_s = {\gamma_F^3+2\Delta^2 \mu \over \pi^2}.
\end{eqnarray}
Since we work at zero temperature, the energy density is given by \cite{Lugones:2002va}
\begin{eqnarray}
\epsilon = 3 \mu n_B-P.
\end{eqnarray}
Note that in the limit $m_s \to 0$ the particle densities become equal, and the EoS  takes the simple form $\epsilon = 3P+4B$. 
 Pairing introduces the $\Delta^2$ term in Eq. (\ref{cfl});  thus the EoS picks up an extra term from CFL contribution as $\epsilon = 3 P+ 4B-6\Delta^2 \mu^2 / \pi^2$. However, this situation becomes more complicated and  difficult to obtain an exact expression for an EoS when $m_s \neq 0$.  Thus, the series upto the order $\Delta^2$ and $m_s^2$ helps us to keep the EoS very simple and useful for most calculations, which yield \cite{lugo02}  
\begin{eqnarray}
P &=& {3\mu^4 \over 4\pi^2}+ {9\beta \mu^2 \over 2\pi^2}-B,\\
\epsilon &=& {9\mu^4 \over 4\pi^2}+ {9\beta \mu^2 \over 2\pi^2}+B, 
\end{eqnarray}
where
\begin{eqnarray}
\beta = -\frac{m_s^2}{6}+ \frac{2\Delta^2}{3}.
\end{eqnarray}
From the above expressions we can obtain an analytic
expression for $\epsilon = \epsilon(P)$:
\begin{eqnarray}
\epsilon = 3P+ 4B -{9\beta \over \pi^2} \left\{\left[{4\pi^2 (B+P) \over 3}+9\beta^2 \right]^{1/2}-3\beta \right\}. \nonumber \\\label{eos} 
\end{eqnarray}
Since the case of Eq. (\ref{eos}) contains three free parameters which are $m_s$, $\Delta$ and $B$, respectively. In this work, we choose the values of the color superconducting gap $\Delta \sim 50-150$ ${\rm MeV}$, see Refs. \cite{Alford1998,Rapp1998}. In order to be absolutely
stable the energy per baryon of the CFL phase, the bag constant $B$  always to be greater than 57 MeV/fm$^3$ \cite{far84} and the strange quark mass $m_{s}$ to be $m_s \sim 50-150 \text{ MeV}$. We use all the values for our QS which  fall inside the \textit{stability windows} presented in Ref. \cite{Flores:2017hpb}. 

%%%%%%%%%%%%%%%%%%%%%%%%%%%%%%%

\subsection{The charge density relation}

We also need to specify the charge density as the matter field carries a net electric charge.
As discussed in \cite{Ray:2003gt,Arbanil:2013pua}, we assume that  the charge density is proportional to the energy density, i.e.
\begin{equation}
\rho_{ch} = \alpha \: \epsilon \, ,
\end{equation}
with $\alpha$ being the charge fraction. This is a reasonable assumption in the sense that large mass can hold large amount of charge. Moreover,
\begin{equation}
\frac{Q}{M} = \frac{\rho_{ch}}{\epsilon} = \alpha,
\end{equation}
and since for the RN solution $Q \leq M$, the charge fraction lies in the range of $0 \leq \alpha \leq 1$.

\begin{figure}
    \centering
    \includegraphics[width = 7.8cm]{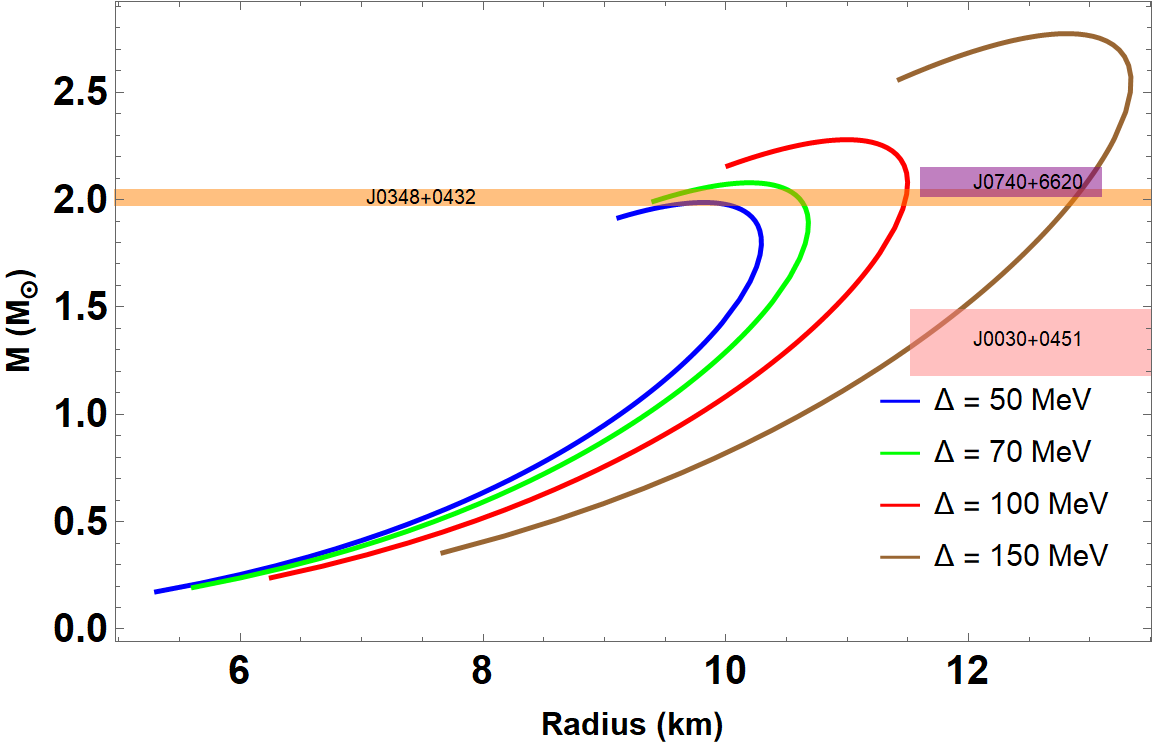}
    \includegraphics[width = 7.8cm]{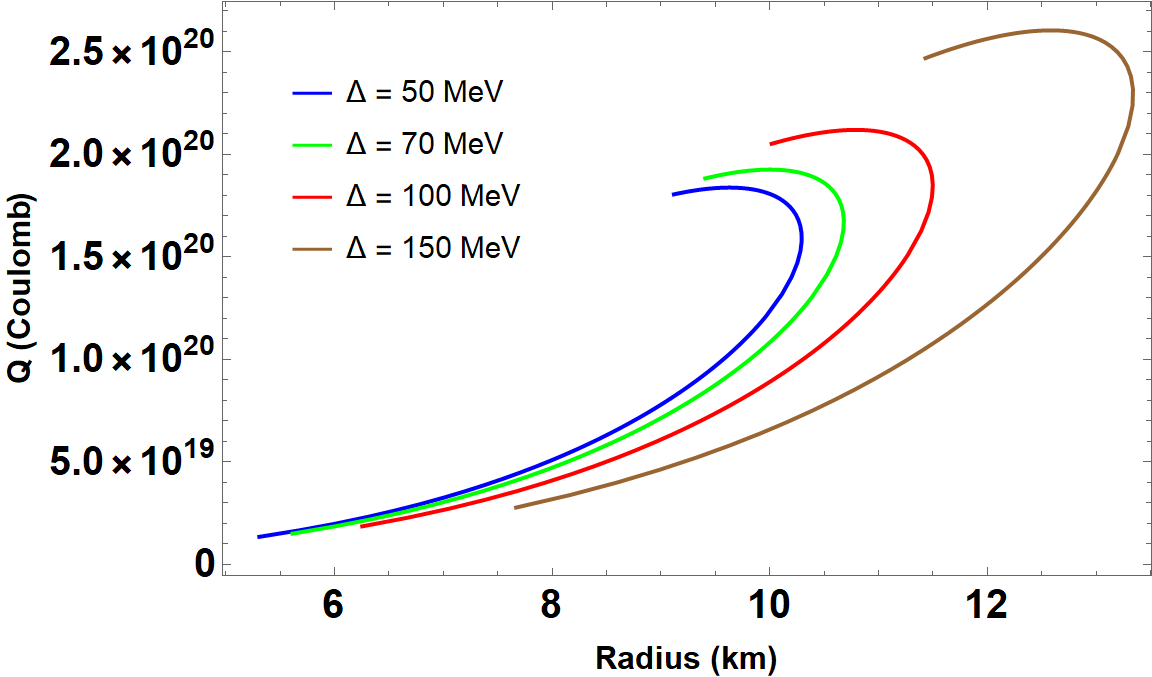}
    \includegraphics[width = 7.8cm]{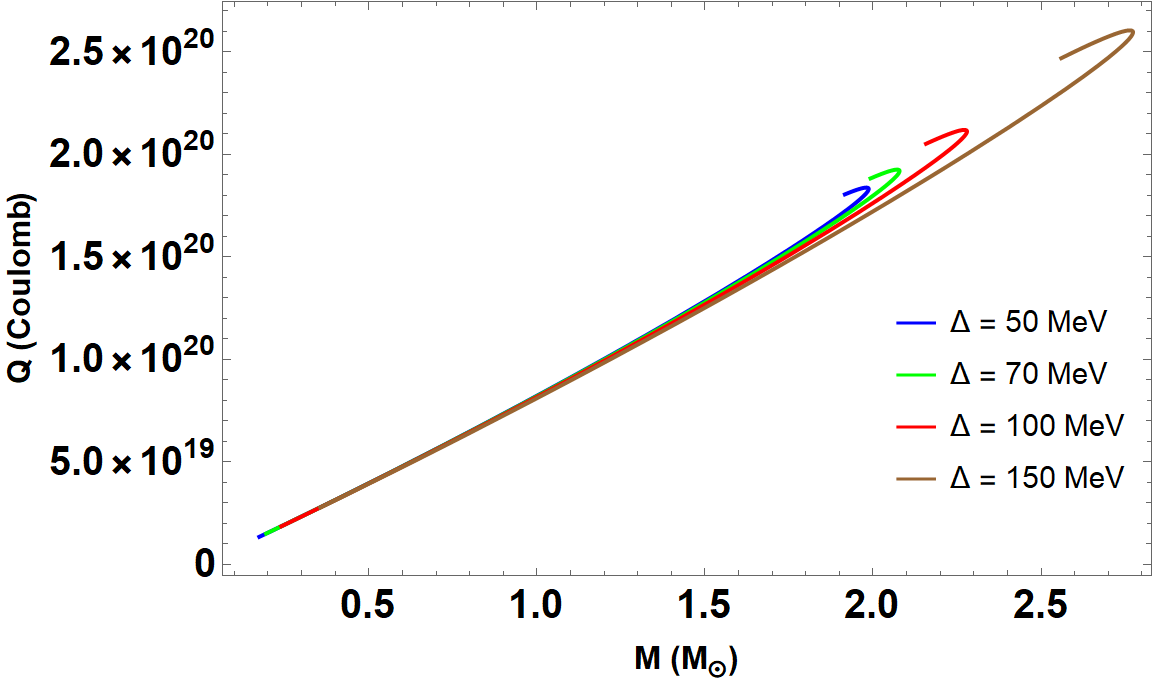}
    \caption{
    The top panel shows the behavior of the mass-radius relation, while the one in the middle shows the variation of the total charge with the radius. The bottom panel shows the variation of the charge with the mass for different $\Delta = 50, 70, 100, 135 \, \text{MeV}$. For plotting we set the other parameters $B = 70 \text{ MeV/fm}^3$, $m_s=150 \,{~\rm MeV}$ and $\alpha = 0.43462$, respectively. The horizontal bands show the observational constraints from various pulsar measurements: PSR J0348+0432 (Orange) \cite{Antoniadis:2013pzd}, PSR J0740+6620  (Purple) \cite{Fonseca:2021wxt} and GW 190814 event with the mass range 2.50-2.67 $M_{\odot}$ (Yellow) \cite{LIGOScientific:2020zkf}, see text for details.}
    
    \label{Result1}
\end{figure}

\begin{figure}
    \centering
    \includegraphics[width = 7.8cm]{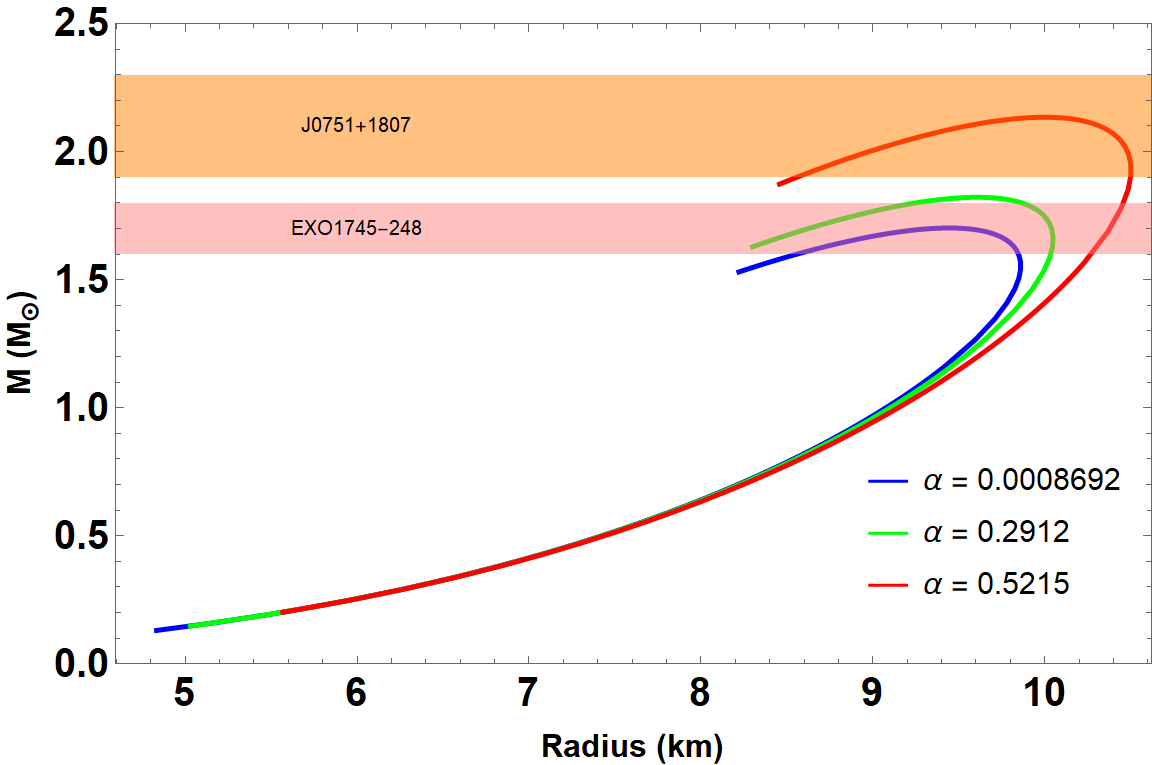}
    \includegraphics[width = 7.8cm]{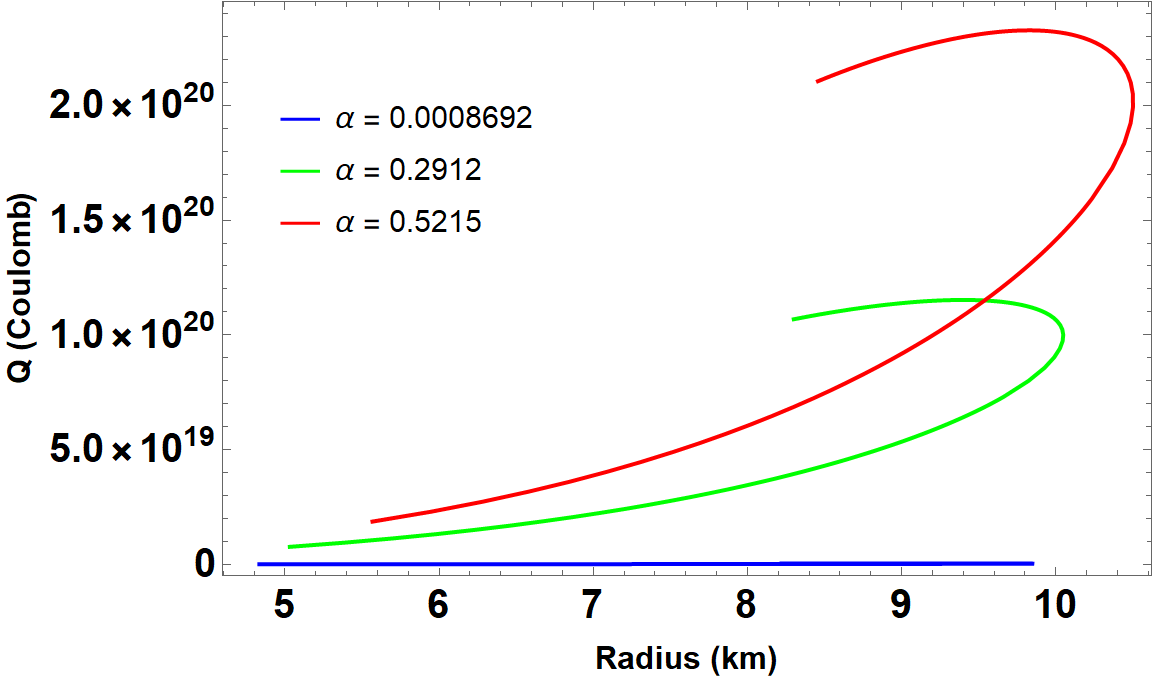}
    \includegraphics[width = 7.8cm]{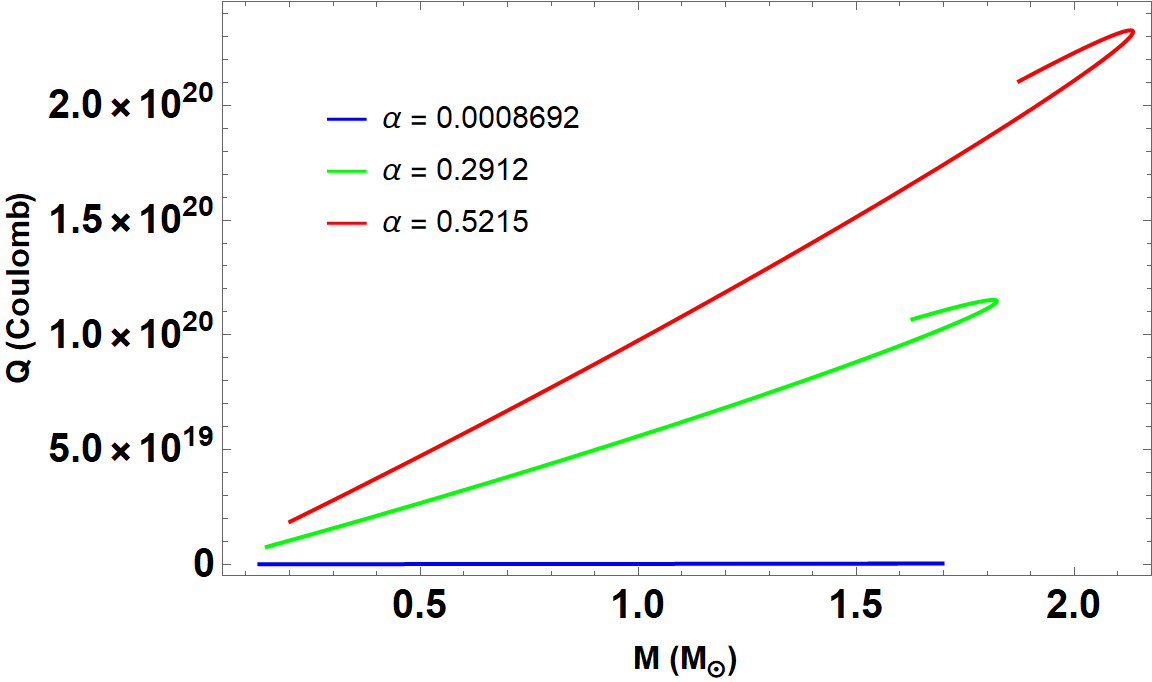}
    \caption{
    We maintain the same sequence as of Fig. \ref{Result1}. The stars described by the EoS (\ref{eos}) with $B = 70 \text{ MeV/fm}^3$, $m_s=150 \,{~\rm MeV}$, $\Delta = 50 \, \text{MeV}$ and different values of $\alpha = 0.0008692, 0.2912, 0.6954$, respectively. The three horizontal bands show the observational constraints from EXO 1745-248 (Pink) \cite{Ozel:2008kb},  PSR J0751+1807 (Orange) \cite{Nice:2005fi} and GW 190814 event with the mass range 2.50-2.67 $M_{\odot}$ (Yellow) \cite{LIGOScientific:2020zkf}.}
    
    \label{Result2}
\end{figure}

\begin{figure}
    \centering
    \includegraphics[width = 7.8cm]{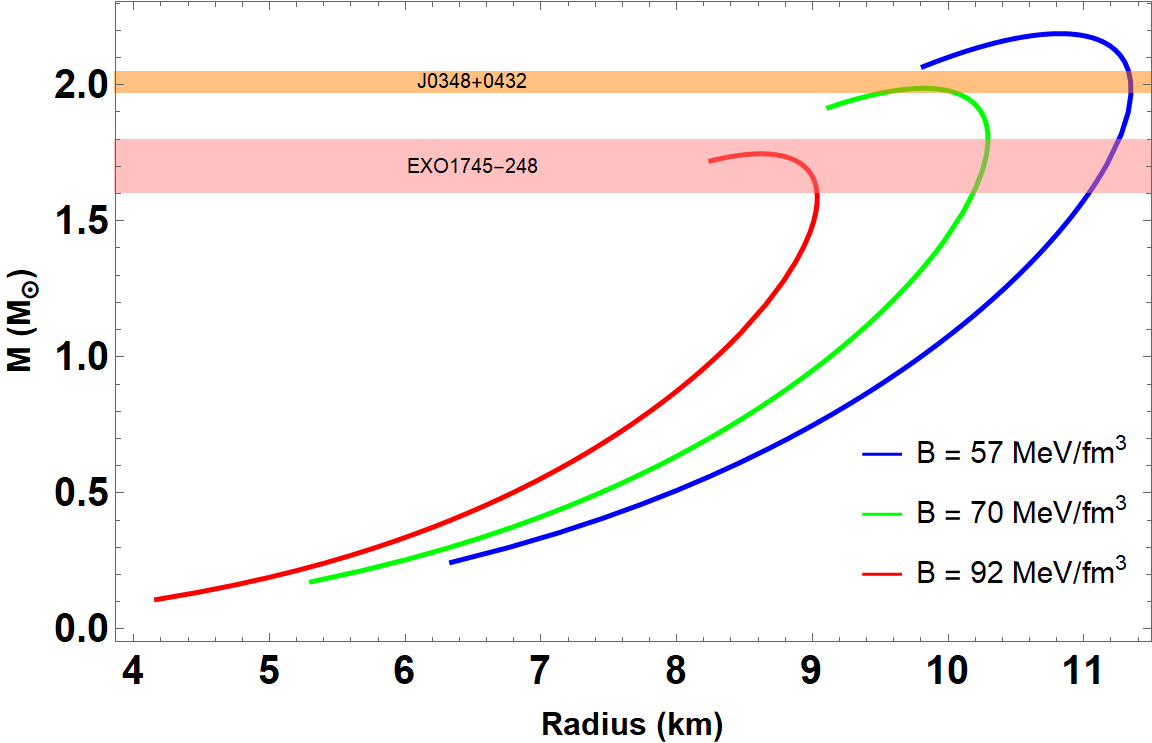}
    \includegraphics[width = 7.8cm]{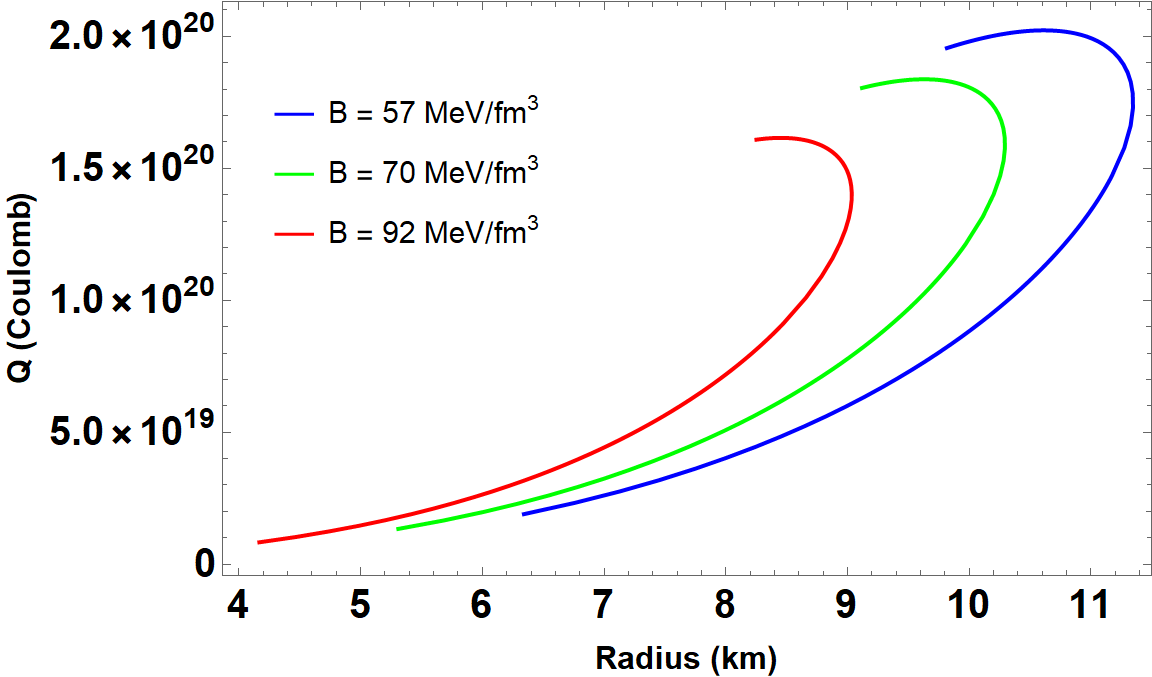}
    \includegraphics[width = 7.8cm]{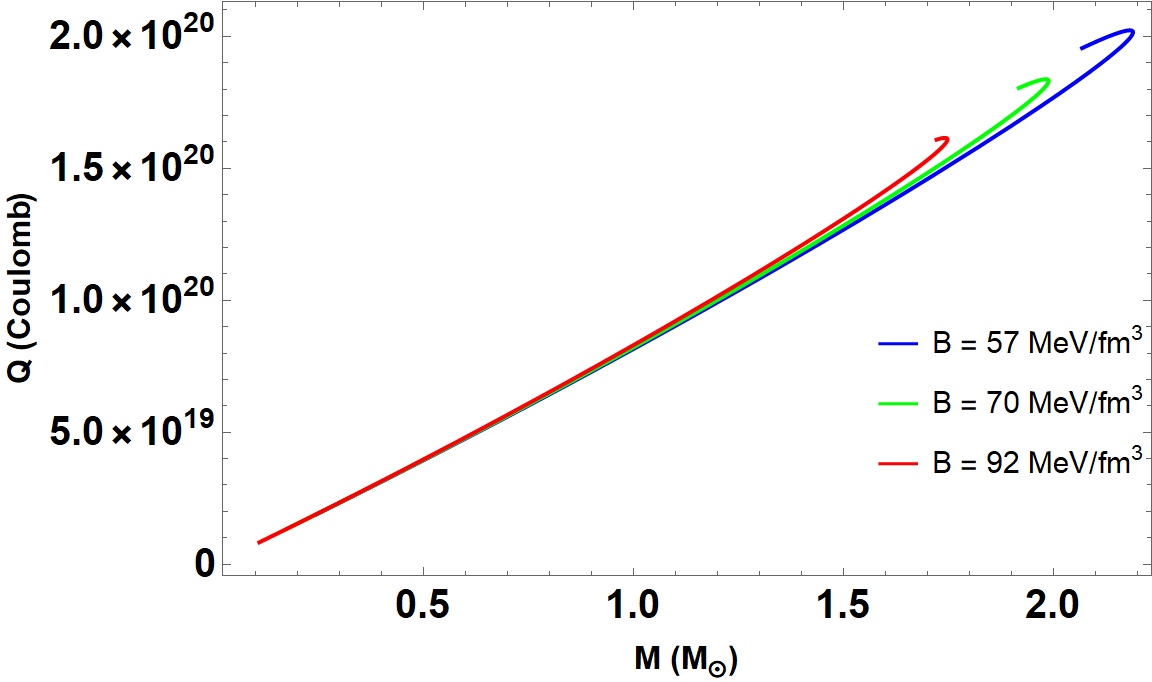}
    \caption{
    We repeat the same sequence as of Fig. \ref{Result1} for the set parameters  $\Delta = 50$ MeV, $m_s = 150$ MeV, $\alpha = 0.43462$ and different values of $B = (57, 70, 92)$ MeV/fm$^3$. The horizontal bands indicate the masses of PSR J0348+0432 \cite{Antoniadis:2013pzd} and  EXO 1745-248  \cite{Ozel:2008kb} pulsars.}
    \label{Result3}
\end{figure}

\begin{figure}
    \centering
    \includegraphics[width = 7.5cm]{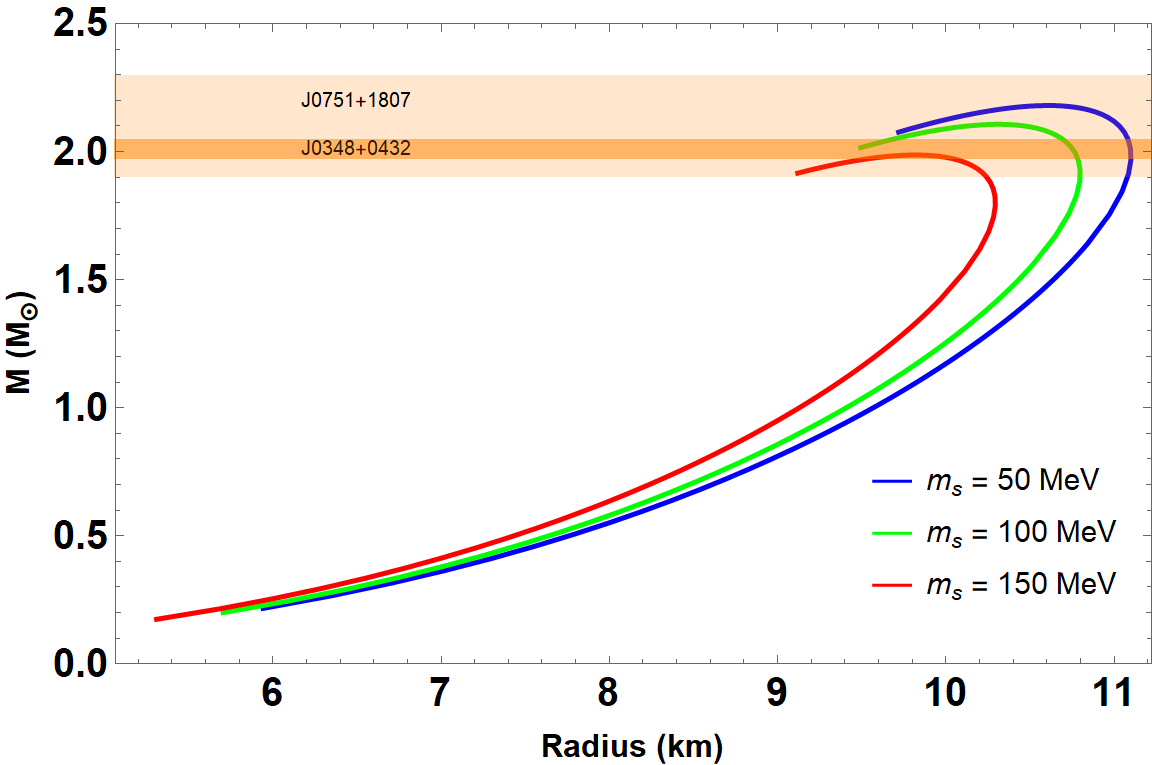}
    \includegraphics[width = 7.5cm]{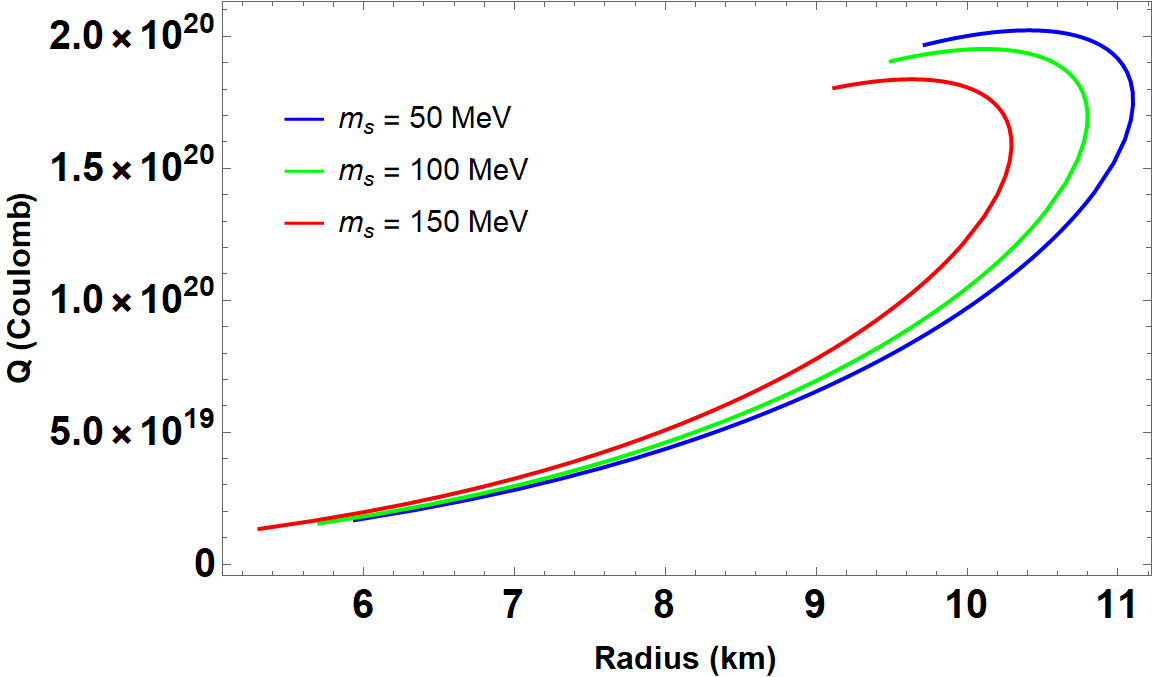}
    \includegraphics[width = 7.5cm]{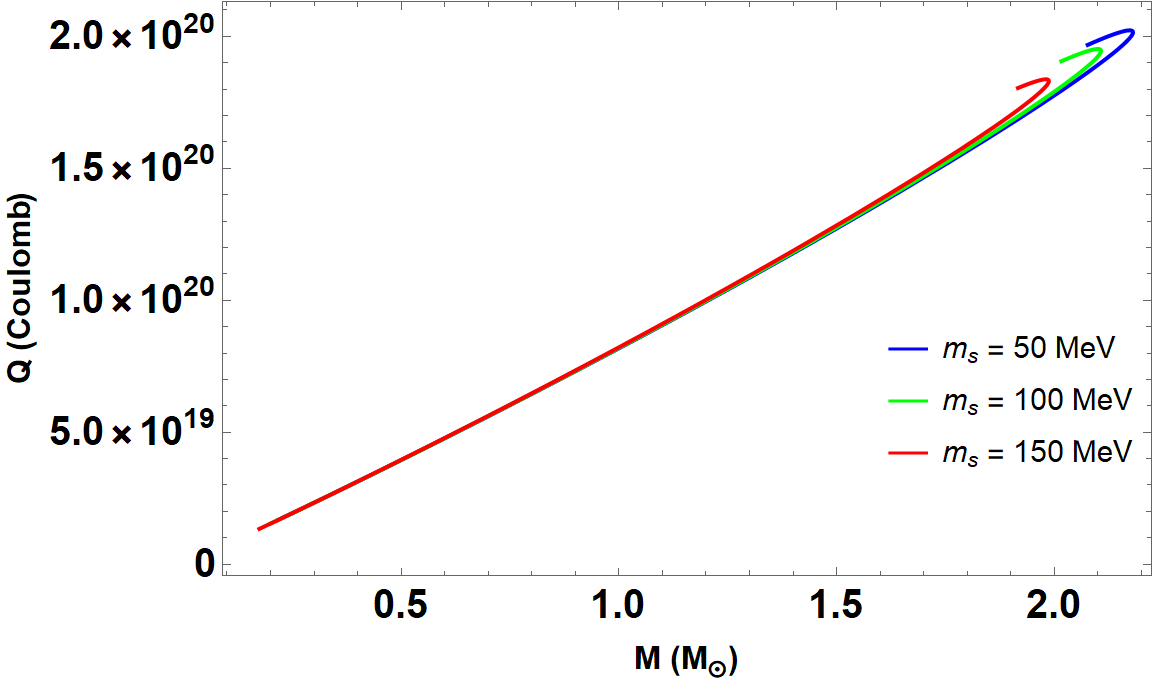}
    \caption{Same as Fig. \ref{Result1}, but for the set of parameters $\Delta = 50$ MeV, $B = 70$ MeV/fm$^3$,  $\alpha = 0.43462$ and different values of $m_s = (50, 100, 150)$ MeV. The horizontal bands indicate two massive pulsars PSR J0751+1807 (Pink) \cite{Nice:2005fi} and PSR J0348+0432 (Orange) \cite{Antoniadis:2013pzd}.}
    \label{Result4}
\end{figure}

%%%%%%%%%%%%%%%%%%%%%%%%%%%%%%%%%%%%%%%%
\section{Numerical results}\label{sec5}
%%%%%%%%%%%%%%%%%%%%%%%%%%%%%%%%%%%%%%%%

In this section we present our main numerical results for the given EoS which was discussed in Sec. \ref{sec4}. To get a good estimate of how quark stars might be, we vary $\Delta$, charge fraction $\alpha$, the bag constant $B$ and the strange quark mass $m_s$, over a physically reasonable range and calculate the mass-radius relations in each case. Here, the mass of the star is measured in units of solar masses $M_{\odot}$, the radius of the stars in $ {\rm Km}$, the total electric charge in $C$ (Coulomb),  while the energy density, the pressure and the bag constant are expressed in $\text{MeV/fm}^3$.  In the following analysis we discuss four sets of solutions (i) variation of color superconducting gap $\Delta = (50-150) \text{ MeV}$, (ii) variation of  $\alpha$ in the range of $\alpha \approx (0-1)$, (iii) variation of bag constant $B = (57-92)$ MeV/fm$^3$, and (iv) variation of the strange quark mass $m_s = (50-150) \text{ MeV}$, respectively.  The results of our calculations are displayed in five distinct 
Tables \ref{tab:table1}-\ref{tab:table5}, respectively.

\subsection{Mass-Radius relation}\label{mr}

\begin{figure}
    \centering
    \includegraphics[width = 7.8cm]{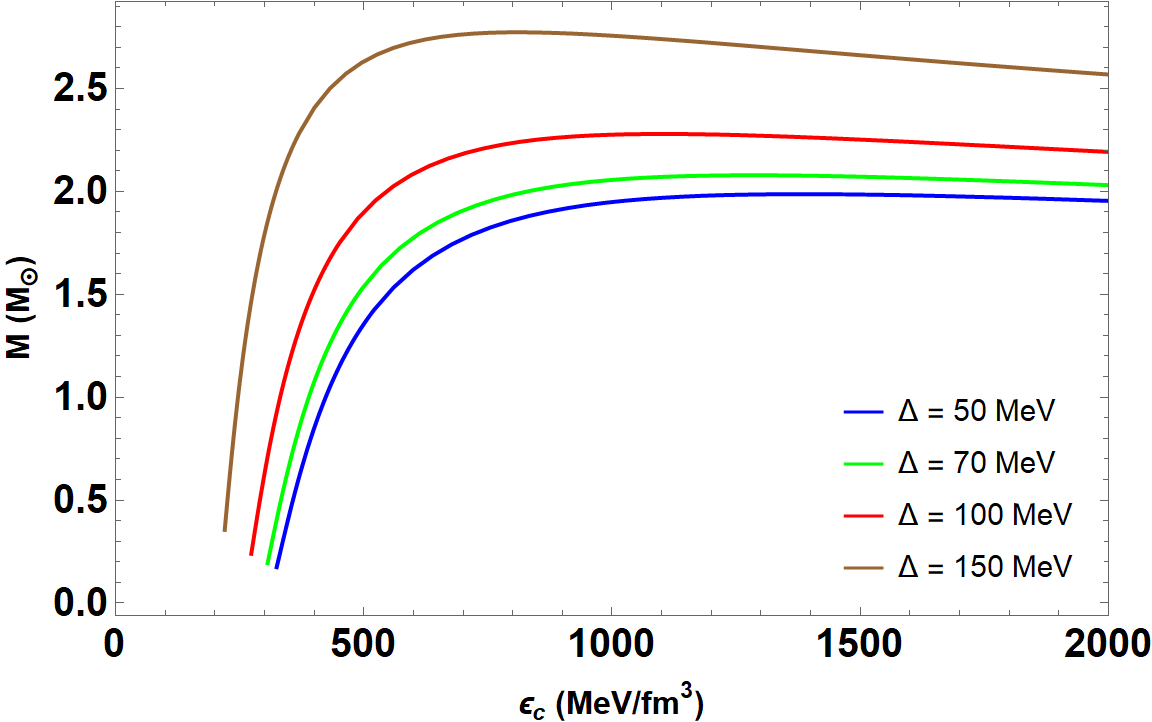}
    \includegraphics[width = 7.8cm]{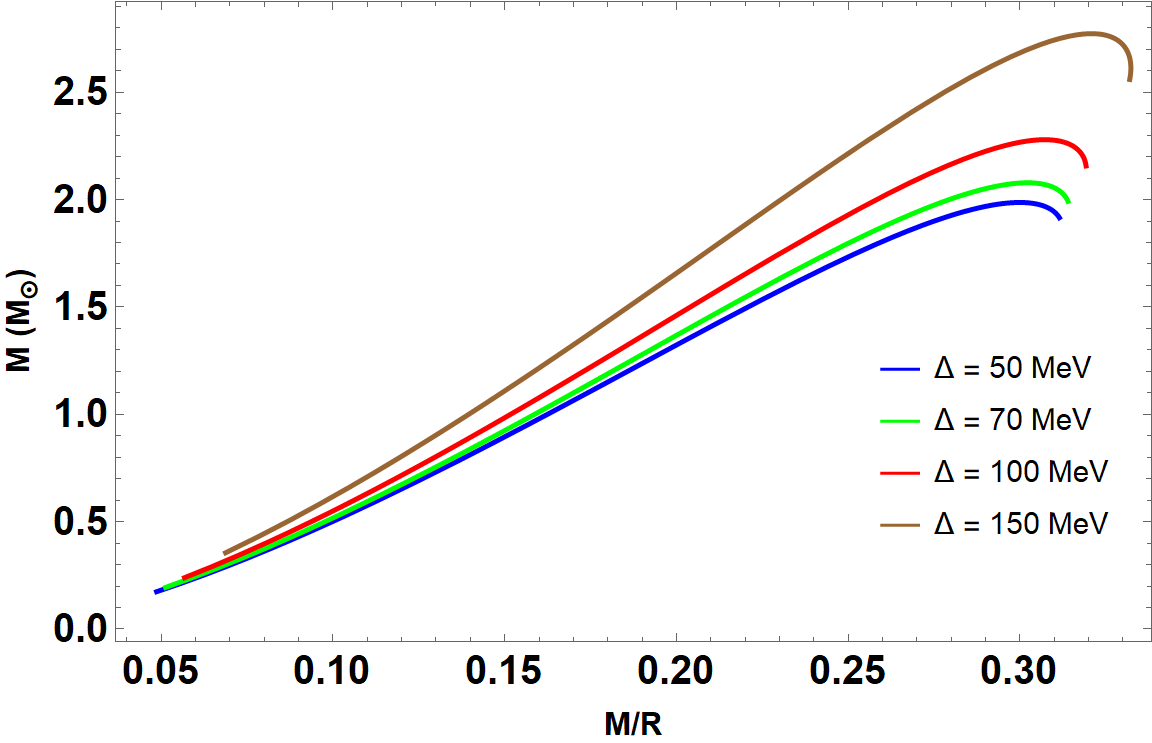}
    \caption{The profiles of mass-energy density and compactness with the same set of parameter in Fig. \ref{Result1}.}
    \label{Result11}
\end{figure}

We have obtained the $M-R$, $Q-R$ and $Q-M$ relations for some values of the central pressure with a fixed value of $\alpha = 0.43462$ and different values of $\Delta = (50, 70, 100, 135) ~\, \text{MeV}$.
The upper panel of Fig. \ref{Result1} shows that the maximum mass of the quark star exceeds the $2 M_{\odot}$ for $\Delta > 50$ $\text{MeV}$. For a given mass the radius increases with $\Delta$, and therefore the factor of compactness, $C=M/R$, decreases with $\Delta$. The same conclusion is reached looking at the lower panel of Fig. \ref{Result11}. Furthermore, it is clear from panels 1 and 3 of Fig. \ref{Result1} that the highest mass and the corresponding radius of the stars increase with $\Delta$, see Table \ref{tab:table1}. This is due to the effect of the repulsive force that a charged star has, and supports the existence of more massive stars avoiding the gravitational collapse. The slope of the $Q-M$ curves come from the different color superconducting gap. In panel 2, we show the $Q-R$ diagram for some values of $\Delta$. Notice that the total charge increases with the total mass until it reaches the point of maximum charge value. With the knowledge of the mass-radius relations predicted by different astrophysical groups, our findings are compatible with the existence of the massive pulsars PSR J0740+6620 having a mass of 2.08$\pm$0.07$M_{\odot}$ with its predicted radius 12.35$\pm$ 0.75 (Purple) \cite{Fonseca:2021wxt}, PSR J0348+0432 with $2.01 \pm 0.04 M_{\odot}$ (Orange) \cite{Antoniadis:2013pzd} and GW 190814 event with the mass range 2.50-2.67 $M_{\odot}$ (Yellow) \cite{LIGOScientific:2020zkf} represented by horizontal stripes in Fig. \ref{Result1}. In Table \ref{tab:table1}, we summarize a few relevant quantities of QSs that are used in the present work and in Fig. \ref{Result1}.

In Fig. \ref{Result2}, we repeat the same sequence as of Fig. \ref{Result1}, but this time we show the impact of $\alpha$ from 0.0008692 to 0.6954 on the properties of the QSs, setting $\Delta = 50$ $\text{MeV}$ and $m_s = 150 \,{~\rm MeV}$. The shape of the profiles $M-R$ and $Q-R$ is qualitatively similar to
those of Fig. \ref{Result1}. Similarly to that figure, the highest mass and the corresponding radius increase with $\alpha$, although the factor of compactness, $M/R$, decreases with $\alpha$, a feature observed elsewhere before, e.g. in  \cite{Arbanil:2013pua}. This model provides best agreement with the pulsars PSR J0751+1807 has a mass of $2.1 \pm 0.2 M_{\odot}$ (Orange) \cite{Nice:2005fi},  EXO 1745-248 with $M=1.7 \pm 0.1 \,M_\odot$ (Pink) \cite{Ozel:2008kb} and GW 190814 event with the mass range 2.50-2.67 $M_{\odot}$ (Yellow) \cite{LIGOScientific:2020zkf}.  Plausible values of some parameters describing QS properties are shown in Table \ref{tab:table2} below. It is evident from the Table \ref{tab:table2} that the maximum mass of QS can excess 2 $M_{\odot}$
if we increase the value of $\alpha$ sufficiently large.

%\textcolor{red}{Now we consider the case of $\Delta = 150$ MeV in Fig. \ref{Result1}, the stars with mass 2.08 $M_{\odot}$ have the radiuses 9.97 and 12.97 km and the central energy density 8795.0 and 332.9 MeV/fm$^3$, respectively. The star with mass 1.60 $M_{\odot}$ has radius 12.20 km and the central energy density 284 MeV/fm$^3$. The star with mass 1.44 $M_{\odot}$ have radius 11.85 km and the central energy density 272.4 MeV/fm$^3$.}

%\textcolor{red}{Let consider the case of $\alpha = 0.695392$ in Fig. \ref{Result2}, the stars with mass 2.08 $M_{\odot}$ have the radiuses 8.25 and 11.00 km and the central energy density 6254.7 and 540.61 MeV/fm$^3$, respectively. The star with mass 1.60 $M_{\odot}$ has radius 10.43 km and the central energy density 432 MeV/fm$^3$. The star with mass 1.44 $M_{\odot}$ have radius 10.15 km and the central energy density 413.5 MeV/fm$^3$.}

Similarly, for plotting the Fig. \ref{Result3}, we take $B$ as a phenomenological model input and choose the following three values of $B = (57, 70, 92)$ MeV/fm$^3$. The other set of parameters are  $\Delta = 50$ MeV, $m_s = 150$ MeV and  $\alpha = 0.43462$, remain fixed. From Fig. \ref{Result3}, we can see that for less interacting quarks the maximum masses and their corresponding maximum radii have larger values.  Moreover, the effect of interacting quarks has a similar trend on the $Q-R$ and $Q-M$ diagrams. In panel 3, the total charge grows with the total mass until it reaches the point of maximum charge value and then decreases with the increasing mass. The maximum mass for $B = 57$ MeV/fm$^3$ is about 2.19 $M_{\odot}$ and the corresponding radius of around 10.84 {\rm Km} (see Table \ref{tab:table3}, for more details). Notice that the model is consistent with stringent constraints on the maximum NS mass coming from PSR J0348+0432 (Orange) \cite{Antoniadis:2013pzd} and EXO 1745-248  (Pink) \cite{Ozel:2008kb}. 

Another important quantity for this discussion is the effect of the strange quark mass $m_s$, which we show in Fig. \ref{Result4}. The diagrams $M-R$, $Q-R$ and $Q-M$ are plotted for three different values of $m_s = (50, 100, 150)$ MeV with $\Delta = 50$ MeV, $B = 70$ MeV/fm$^3$ and  $\alpha = 0.43462$, respectively. The trend of figures are almost similar to that of Fig. \ref{Result3}. We have allowed the range of values of $m_s = 150$ MeV that fall inside the stability windows to yield stars of maximum mass up to $M= 2.13 \,M_\odot$. The horizontal strips represent the observational constraints, from top to bottom the first two delimit the band from $(1.99-2.19)~M_{\odot}$, which contains masses of the super-massive pulsars PSR J0751+1807 (Pink) \cite{Nice:2005fi} and PSR J0348+0432 (Orange) \cite{Antoniadis:2013pzd}.   Table \ref{tab:table4} summarizes  the maximum mass, total charge, radius and the compactness of the charged star for different values of strange quark mass $m_s$. 

Finally, in Table \ref{tab:table5}, we summarize the main aspects of the models used in this work
i.e., maximum mass constraints for 1.4$M_{\odot}$, 1.6$M_{\odot}$ and  2.08$M_{\odot}$ considered as a most 
sensational masses, which is used here to probe the sensitivity of our results to the constraint from observations. In particular, the CFL EoS is consistent with the $\sim 2 M_{\odot}$ limit and predicts a radius of
$R_{1.6} = 10.43$ {\rm Km} for a $1.6 M_{\odot}$ neutron star, which is very closed to the observation reported in \cite{Bauswein:2019juq}.

%%%%%%%%%%%%%%FIGURES%%%%%%%%%%%%%%%%%%%%%%%%%%%%%%%%%%%%%%

\begin{figure}
    \centering
    \includegraphics[width = 7.8cm]{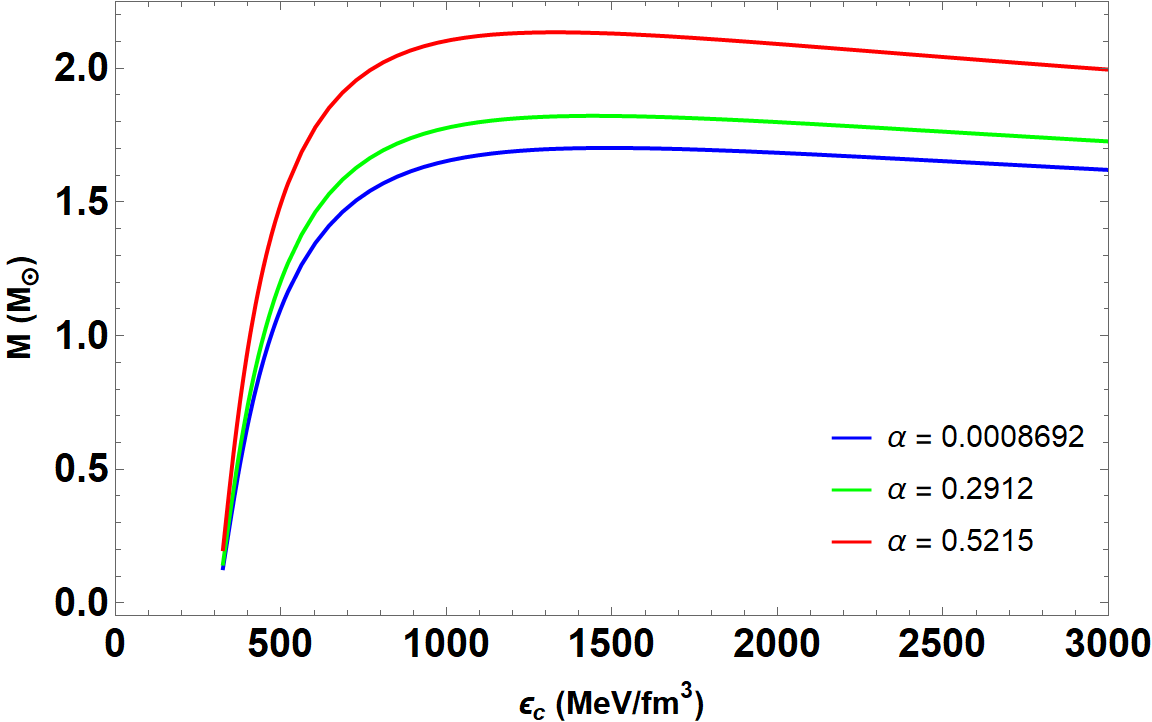}
    \includegraphics[width = 7.8cm]{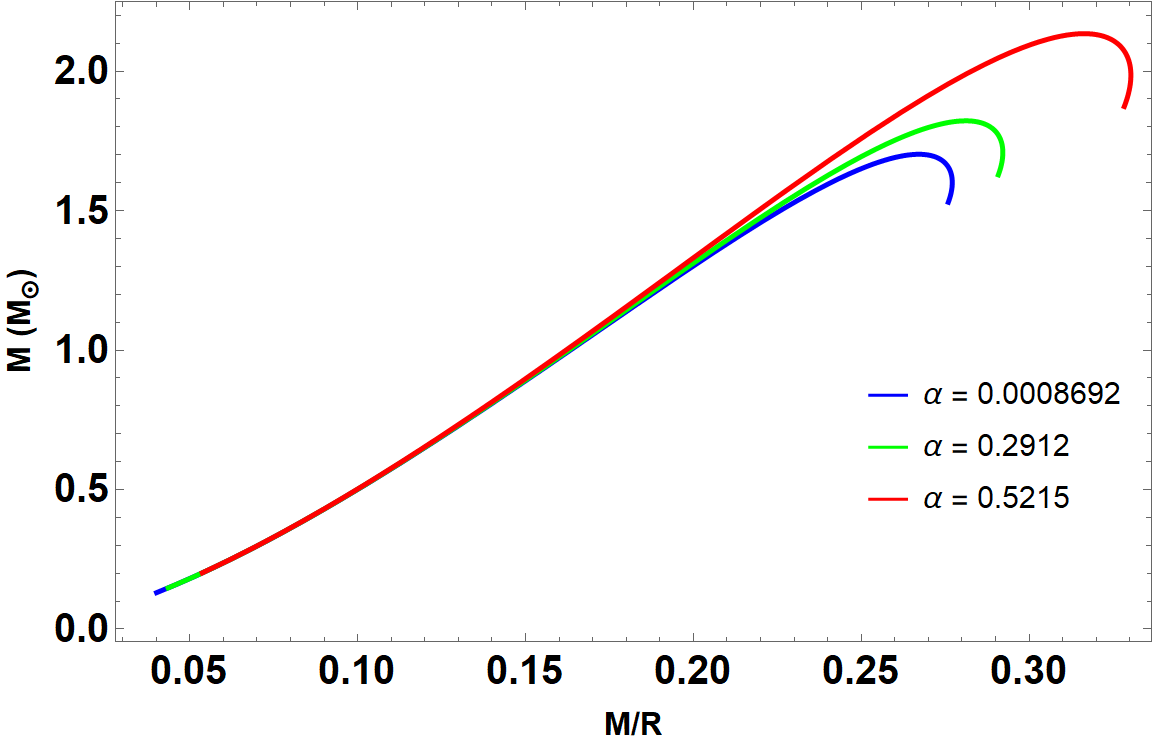}
    \caption{The profiles of mass-energy density and compactness with the same set of parameter in Fig. \ref{Result2}.}
    \label{Result22}
\end{figure}

\begin{figure}
    \centering
    \includegraphics[width = 7.8cm]{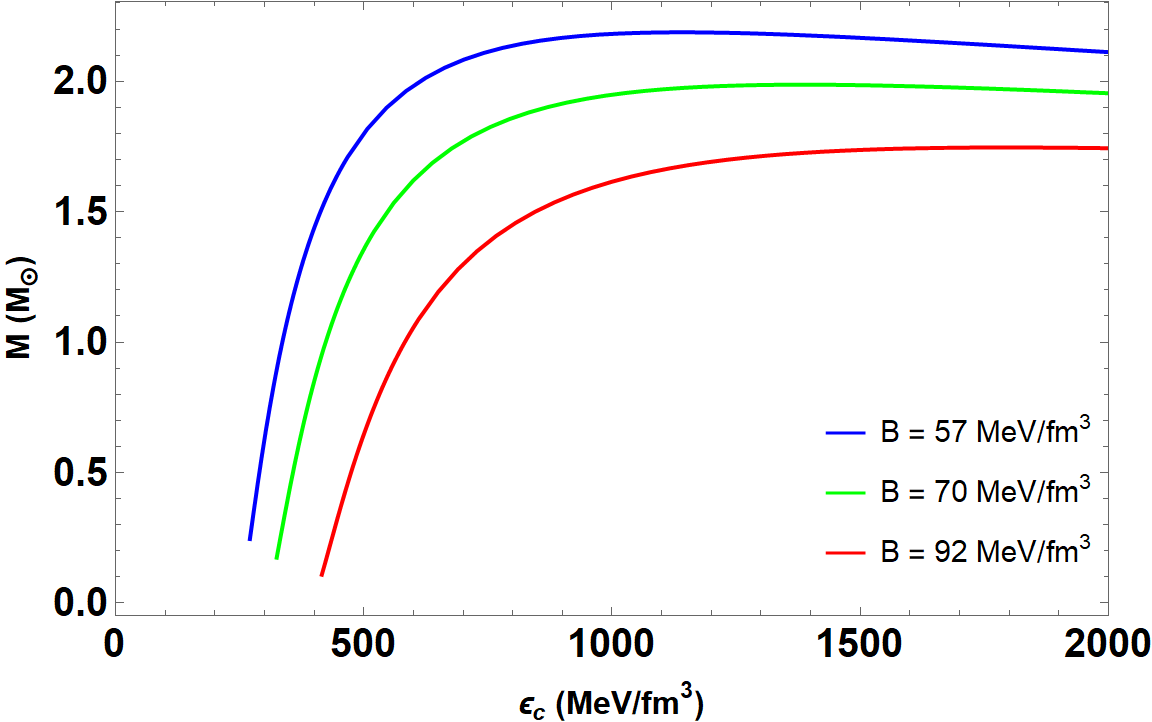}
    \includegraphics[width = 7.8cm]{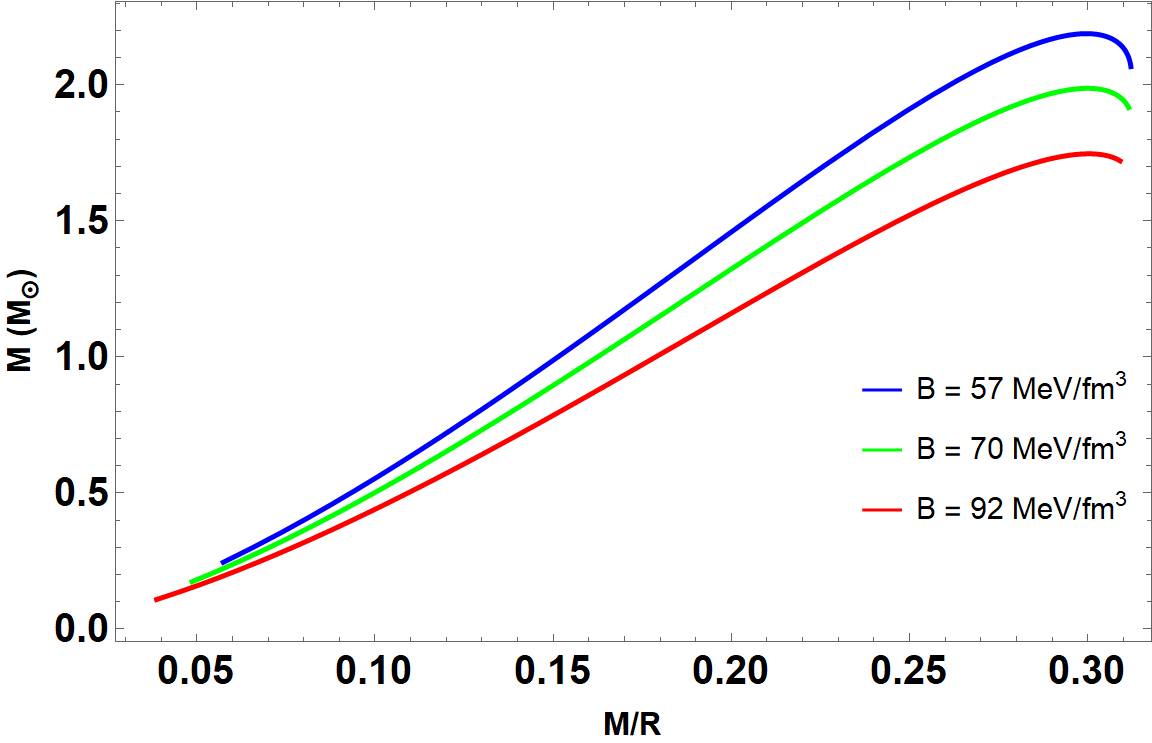}
    \caption{The profiles of mass-energy density and compactness with the same set of parameters in Fig. \ref{Result3}.}
    \label{Result33}
\end{figure}

\begin{figure}
    \centering
    \includegraphics[width = 7.8cm]{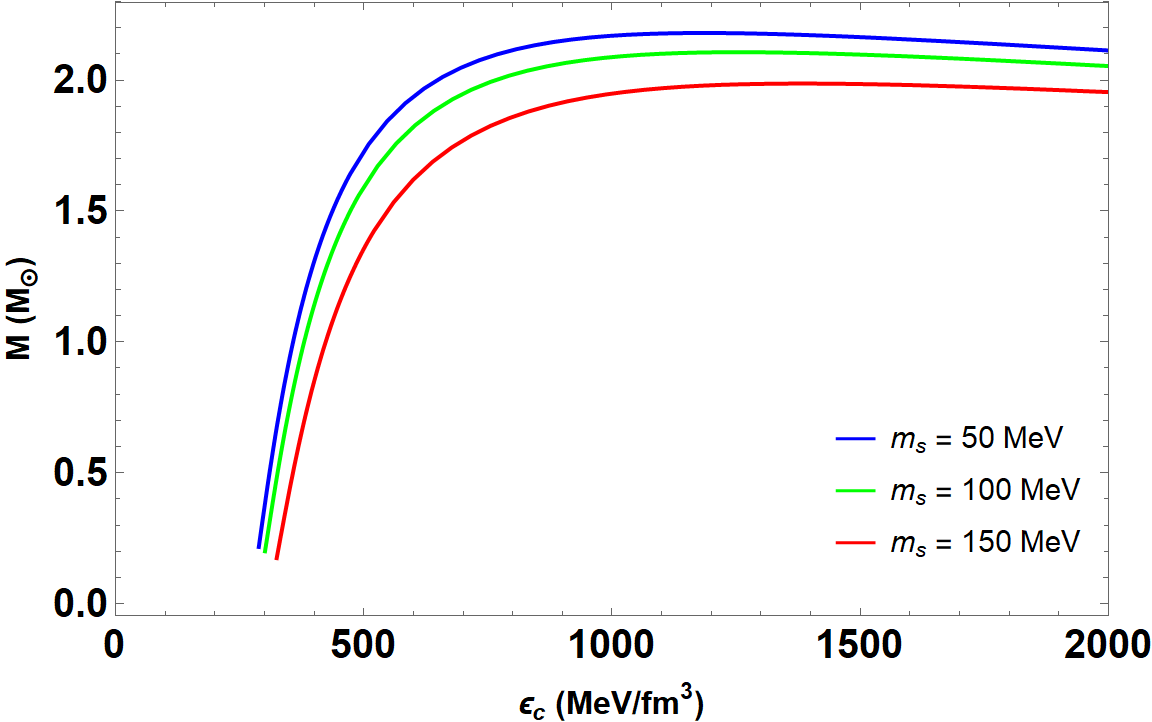}
    \includegraphics[width = 7.8cm]{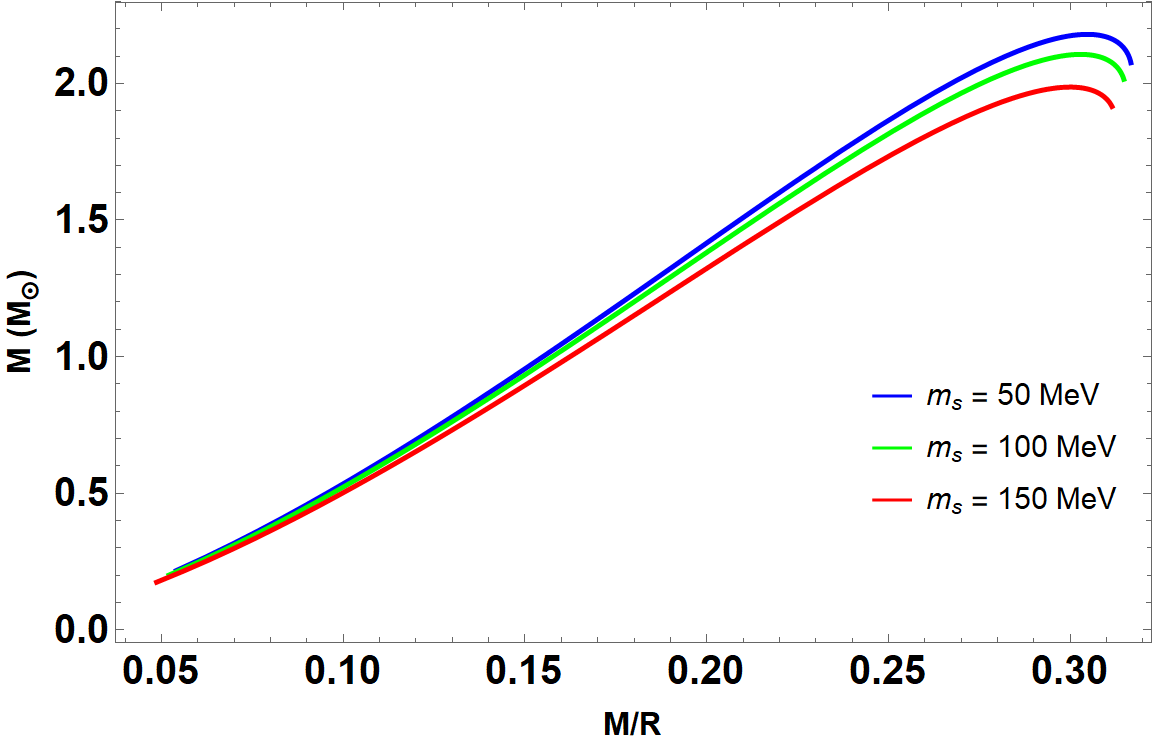}
    \caption{The profiles of mass-energy density and compactness with the same set of parameter in Fig. \ref{Result4}.}
    \label{Result44}
\end{figure}

%%%%%%%%%%%%%%%%%%%%%%%%%%%%%%%%%%%%%%%%%%

\subsection{Dynamical Stability and Compactness} 

We now proceed to study the stellar mass $M$ against the central energy density $\epsilon_c$ for the four different sets of parameters as indicated in Section \ref{mr}. To serve our purpose, we consider the so called static stability criterion \cite{harrison,ZN}
\begin{eqnarray}
&&\frac{d M}{d \epsilon_c} < 0 ~~~ \rightarrow \text{unstable configuration} \\
&&\frac{d M}{d \epsilon_c} > 0 ~~~ \rightarrow \text{stable configuration},
\end{eqnarray}
to be satisfied by all stellar configurations. Note that this is a necessary condition but not sufficient. The mass of the star reaches a maximum value at $\epsilon_c^*$, which grows with the charge fraction in agreement with the $M-R$ profile in Figs. \ref{Result1}-\ref{Result4}, respectively. The results have been shown in Figs. \ref{Result11}, \ref{Result22}, \ref{Result33} and \ref{Result44} (in the top panel), where we plot $M$ (in solar masses $M_{\odot}$) against $\epsilon_c$ for QSs. The variation of central energy densities between the range  $400 \leq \epsilon_c \leq 2000 \,{\rm MeV}/{\rm fm}^{3}$. Therefore, the extremum point of the curve separates the stable from the unstable configuration.

Besides, we also analyze the compactness $M/R$ for charged objects. The value of the ratio $M/R$ has been shown against the maximum mass in the lower panel of Figs. \ref{Result11}-\ref{Result44}, respectively. The main feature to note is that $M/R$ decreases with increasing values of $\Delta$ and $\alpha$, but $M/R$ increases with increasing values $B$ and $m_s$. Finally, we present some of the values of stellar compactness in Tables \ref{tab:table1}-\ref{tab:table5}. In this prescription of quark stars we found that the compactness corresponding to the highest mass increases with $\Delta,\alpha,B$ (see Tables \ref{tab:table1}, \ref{tab:table2} and \ref{tab:table3} for more details), and decreases with $m_s$ (see Tables \ref{tab:table4} for more details).

In Ref. \cite{Buchdahl}, Buchdahl has also obtained the compactness limit for a perfect fluid interior given by $\frac{M}{R} \leq  4/9$ for an object of mass $M$ and radius $R$. The generalization of 
Buchdahl's bound was given for charged gravitational objects by Andr\'{e}asson \cite{Andreasson:2008xw}. The limit was obtained by assuming the strong energy condition, $p_r + 2p_{\perp} \leq \rho$, where $p_r$ and $p_{\perp}$ are the radial and tangential pressures and $\rho$ is the matter energy density, respectively. The Buchdahl-Andr\'{e}asson bound is found to be \cite{Andreasson:2008xw,Chakraborty:2020ifg} 
\begin{eqnarray}\label{bh}
\frac{M}{R} \leq \frac{8}{9}\left(1+\sqrt{1-\frac{8 Q^2}{9 M^2}}\right)^{-1}.
\end{eqnarray}  
When $Q=0$, Eq. (\ref{bh}) gives the usual Buchdahl bound for electrically neutral objects, $M/R \leq 4/9$. On the other hand, when $Q=M$ we obtain for the factor of compactness the limit $M/R \leq 2/3$. For any other value of $Q/M$ between the two extremal cases, the bound on $M/R$ may be seen in the left panel of Figure~1 of \cite{Chakraborty:2020ifg}.
It is clear from the Tables \ref{tab:table1}-\ref{tab:table5} that the star obeys the mass-radius ratio,  which is a requirement of the dynamically stable horizonless charged compact object.

%%%%%%%%%%%%%%%%%%%%%TABLES%%%%%%%%%%%%%%%%%%%%%%%%%%%%%

%%%%%%%%%%%%%%%%%%%%%1st%%%%%%%%%%%%%%%%%%%%

\begin{table}[h]
  \caption{\label{tab:table1} Characteristics of a set of configurations for EoS (\ref{eos}). We list our choices for the parameters which according to the discussion given in Fig. \ref{Result1}. 	The total charge $Q$ of the stars is measured in Coulomb (C).}
\begin{ruledtabular}
\begin{tabular}{ccccccc}
$\Delta$ & $M$ & $R$ & $Q$ & $\epsilon_c$ & $\frac{M}{ R}$\\
${\rm MeV}$ & ($M_{\odot}$) & (Km) & ($\times 10^{20}C$)& $\frac{{\rm MeV}}{{\rm fm^3}}$ & \\
\hline
    50 &  1.99  & 9.83 & 1.83 & 1374 & 0.300\\
    70 &  2.08  & 10.19 & 1.92 & 1295 & 0.302\\
    100 &  2.28  & 11.00 & 2.11 & 1106 & 0.307\\
    135 &  2.61  & 12.21 & 2.43 & 898 & 0.316\\
\end{tabular}
\end{ruledtabular}
\end{table}

%%%%%%%%%%%%%%%%%%%2nd%%%%%%%%%%%%%%%%%%%%%%

\begin{table}[h]
  \caption{\label{tab:table2} Quark star properties using the EoS (\ref{eos}) and their corresponding central energy densities 
	for different values of $\alpha$. We list our choices for the parameters $B$, $m_s$ and $\Delta$, which according to the discussion given in Fig. \ref{Result2}.}
\begin{ruledtabular}
\begin{tabular}{ccccccc}
$\alpha$ & $M$ & $R$ & $Q$ & $\epsilon_c$ & $\frac{ M}{ R}$&  \\
 & ($M_{\odot}$) & (km) & ($\times 10^{20}C$) & $\frac{{\rm MeV}}{{\rm fm^3}}$ &  \\
\hline
    %0.000869 &  1.50  & 8.29 & 0.00285 & 1925 & 0.180\\
    %0.435 &  1.99  & 9.83 & 1.830 & 1.71 & 1374 & 0.3007\\
    %0.291 &  1.60  & 8.43 & 1.01 & 1886 & 0.190\\
    %0.623 &  2.08  & 9.01 & 2.64 & 1617 & 0.231\\
    %0.695 &  2.59  & 10.5 & 3.580 & 2.93 & 1181 & 0.3666\\
    %0.869 &  2.96  & 9.92 & 4.78 & 1193 & 0.444\\
    0.000869 &  1.70  & 9.43 & 0.00325 & 1502 & 0.267\\
    0.435 &  1.99  & 9.82 & 1.83 & 1380 & 0.300\\
    0.695 &  2.59  & 10.50 & 3.59 & 1177 & 0.365\\
\end{tabular}
\end{ruledtabular}
\end{table}

%%%%%%%%%%%%%%%%%%%%3rd%%%%%%%%%%%%%%%%%%%%%

\begin{table}[h]
  \caption{\label{tab:table3} Based on the parameters of Fig. \ref{Result3}, we enlist the maximum-mass, total charge, radius and the compactness  of the charged star with their corresponding values of central energy density.}
\begin{ruledtabular}
\begin{tabular}{cccccc}
$B$ & $M$ & $R$ & $Q$ & $\epsilon_c$ & $\frac{ M}{ R}$ \\
$\frac{{\rm MeV}}{{\rm fm^3}}$ & ($M_{\odot}$) & (Km)  &($\times 10^{20}C$)& $\frac{{\rm MeV}}{{\rm fm^3}}$ & \\
\hline
 57 &  2.19  & 10.84 & 2.01 & 1128 & 0.299\\
 70 &  1.99  & 9.83 & 1.83 & 1374 & 0.300\\
 92 & 1.75  & 8.61 & 1.61 & 1810 & 0.301\\
\end{tabular}
\end{ruledtabular}
\end{table}

%%%%%%%%%%%%%%%%%%%%4th%%%%%%%%%%%%%%%%%%%%%

\begin{table}[h]
  \caption{\label{tab:table4}  Based on the parameters of Fig. \ref{Result4}, we enlist the maximum-mass, total charge, radius and the compactness  of the charged star with their corresponding values of central energy density.}
\begin{ruledtabular}
\begin{tabular}{cccccc}
$m_s$ & $M$ & $R$ & $Q$ & $\epsilon_c$ & $\frac{ M}{ R}$\\
 ${\rm MeV}$ & ($M_{\odot}$) & (Km)  &($\times 10^{20}C$)& $\frac{{\rm MeV}}{{\rm fm^3}}$ & \\
\hline
  50 &  2.19  & 10.61 & 2.01 & 1180 & 0.305\\
 100 &  2.11  & 10.32 & 1.94 & 1246 & 0.303\\
 150 &  1.99  & 9.83 & 1.83 & 1374 & 0.300\\
\end{tabular}
\end{ruledtabular}
\end{table}

%%%%%%%%%%%%%%%%%%%%5th%%%%%%%%%%%%%%%%%%%%%

\begin{table}[h]
  \caption{\label{tab:table5}   Maximum mass constraints for 1.44$M_{\odot}$, 1.6$M_{\odot}$ and  2.08$M_{\odot}$ based on the choice for the parameters $\Delta = 50$ MeV, $\alpha = 0.695392 \, \text{MeV}$, $B = 70$ ${\rm MeV}/{\rm fm}^{3}$ and $m_s = 150$ MeV,  respectively.}
\begin{ruledtabular}
\begin{tabular}{cccccc}
$M$ & $R$ & $Q$ & $\epsilon_c$ & $\frac{ M}{ R}$\\
  ($M_{\odot}$) & (Km)  &($\times 10^{20}C$)& $\frac{{\rm MeV}}{{\rm fm^3}}$ & \\
\hline
    1.40  & 10.12 & 1.80 & 412 & 0.206\\
    1.60  & 10.43 & 2.08 & 432 & 0.227\\
    2.08  & 10.99 & 2.76 & 529 & 0.280\\
\end{tabular}
\end{ruledtabular}
\end{table}
%%%%%%%%%%%%%%%%%%%%%%%%%%%%%%%%%%%%%%%%%

%%%%%%%%%%%%%%%%%%%%6th%%%%%%%%%%%%%%%%%%%%%

\iffalse
\begin{table}[h]
  \caption{\label{tab:table6} The obtained values of the maximum-mass, radius and their corresponding central energy density for several different values of $\Delta$ with a fixed value of charge fraction $\alpha = 0.086924 \, \text{MeV}$.}
\begin{ruledtabular}
\begin{tabular}{cccccc}
$\Delta$ & $M$ & $R$ & $Q$ & $\epsilon_c$ & $\frac{ M}{ R}$\\
\\
\hline
    48 &  1.40  & 7.96 & 0.265 & 2099 & 0.2619\\
    95 &  1.60  & 8.93 & 0.304  & 1672 & 0.2666\\
    152 &  2.08  & 11.11 & 0.400 & 1064 & 0.2786\\
\end{tabular}
\end{ruledtabular}
\end{table}
\fi

%%%%%%%%%%%%%%%%%%%%%%%%%%%%%%%%%%
\section{Conclusions}\label{sec6}
%%%%%%%%%%%%%%%%%%%%%%%%%%%%%%%%%%

We have studied electrically charged strange quark stars in the context of Einstein-Maxwell theory. The matter field is assumed to be color-flavor-locked (CFL) phase of color superconductivity, while the charge density is proportional to the mass density. The quark matter in the CFL phase may be the true ground state of hadronic matter with charge neutrality. Using the EoS we have solved the Tolman-Oppenheimer-Volkoff (TOV) equations, which describe hydrostatic equilibrium, together with the appropriate boundary conditions. The Reissner-Nordstr\"{o}m spacetime has been considered to match the interior solution at the surface of the star. Finally, we explore the compatibility of QSs in the CFL phase with a set of maximum mass constraints corresponding to PSR J0740+6620,  PSR J0348+0432,  EXO 1745-248, PSR J0751+1807,  PSR J0030+0451 and GW 190814 event with the mass range 2.50-2.67 $M_{\odot}$.

We have integrated numerically the structure equations for charged spheres, and we have explored the mass-radius $(M-R)$ relation as well as some other physical properties of strange quark stars. The model adopted here is characterized by three parameters that enter into the EoS, $(B,~\Delta,~m_s)$, and also by the charge fraction, $\alpha$, that lies in the range of $0 \leq \alpha \leq 1$.  In particular, here we have explored the $(M-R)$ profile based on the variation of all four free parameters of the model. Our findings confirm the requirement that the EoS supports masses above $M_{\text{max}}> 2 M_{\odot}$ depending on the choice of parameters, see Figs. \ref{Result1}-\ref{Result4}. 
This is in agreement with the results obtained in previous similar works, where it has been observed that charged stars are more massive than their neutral counterparts.
%This is expected, since the presence of a net electric charge acts like an %anisotropic factor leading to more massive stars.}

Also, we have demonstrated the $M-\epsilon_c$ relation and compactness separately. By means of static stability criterion we infer the conditions $\frac{\partial M(\epsilon_c)}{\partial \epsilon_c}$ $\lessgtr 0$ that separate the stable configuration region from the unstable one at the point $(M_{\text{max}}, R_{M_{\text{max}}})$.
Another interesting result in this discussion is the Buchdahl-Andr\'{e}asson bounds on the mass-radius ratio of charged gravitational objects. Our results show that the Buchdahl-Andr\'{e}asson bound is not saturated.  As a result, we see that the proposed model is in good agreement with the stability conditions. Summarizing the results obtained in the present work, within Einstein-Maxwell theory, quark matter in the CFL phase can be potentially used to model stellar structure of relativistic compact objects.

%%%%%%%%%%%%%%%%%%%%%%%%%%%%%%%%%%%%%%%%%%%%%%%%%%%%%%%%
\section*{Acknowlegements}

A. Pradhan thanks to IUCCA, Pune, India for providing facilities under associateship programmes.
 The author G.~P. thanks the Fun\-da\c c\~ao para a Ci\^encia e Tecnologia (FCT), Portugal, for the financial support to the Center for Astrophysics and Gravitation-CENTRA, Instituto Superior T\'ecnico, Universidade de Lisboa, through the Project No.~UIDB/00099/2020 and No.~PTDC/FIS-AST/28920/2017. 
%%%%%%%%%%%%%%%%%%%%%%%%%%%%%%%%%%%%%%%%%%%%%%%%%%%%%%%%

%%%% .......................... BIBLIOGRAPHY .....................

%\bibliography{rlist}

%\bibliographystyle{unsrt}

\end{document}